\documentclass[iop,apj,tighten]{emulateapj}
\usepackage{apjfonts} 
\usepackage{amsmath,amstext}
\usepackage[usenames]{xcolor}
\usepackage[breaklinks,colorlinks,citecolor=blue,linkcolor=magenta]{hyperref}
\usepackage{tabularx,graphicx}
\usepackage[all]{hypcap} 

\newcommand{\revb}[1]{\textcolor{black}{#1}}
\newcommand{\rev}[1]{\textcolor{black}{#1}}
\newcommand{\Msun}{\,M_{\odot}}
\newcommand{\Mstar}{M_{\star}}
\newcommand{\Mnsc}{M_{\mathrm{NSC}}}
\newcommand{\Rnsc}{R_{\mathrm{NSC}}}
\newcommand{\Rh}{R_{\mathrm{h}}}
\newcommand{\epsff}{\epsilon_{\mathrm{ff}}}
\newcommand{\feh}{\mathrm{[Fe/H]}}

\begin{document}

\slugcomment{Submitted to ApJ}
\shorttitle{Nuclear Star Clusters in Cosmological Simulations}
\shortauthors{Brown et al.}

\title{Nuclear Star Clusters in Cosmological Simulations}

\author{Gillen Brown$^{1* \href{https://orcid.org/0000-0002-9114-5197}{\includegraphics[scale=0.4]{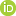}}}$,
Oleg Y. Gnedin$^{1 \href{https://orcid.org/0000-0001-9852-9954}{\includegraphics[scale=0.4]{orcid.png}}}$ and
Hui Li$^{1,2 \href{https://orcid.org/0000-0002-1253-2763}{\includegraphics[scale=0.4]{orcid.png}}}$
}

\altaffiltext{1}{Department of Astronomy, University of Michigan, Ann Arbor, MI 48109, USA}
\altaffiltext{2}{Department of Physics, Kavli Institute for Astrophysics and Space Research, Massachusetts Institute of Technology, Cambridge, MA 02139, USA}
\altaffiltext{*}{gillenb@umich.edu}

\date{\today}

\begin{abstract}
\rev{We investigate the possible connection between the most massive globular clusters, such as $\omega$ Cen and M54, and nuclear star clusters of dwarf galaxies that exhibit similar spreads in age and metallicity. We examine galactic nuclei in cosmological galaxy formation simulations \revb{at $z \approx 1.5$} to explore whether their age and metallicity spreads could explain these massive globular clusters.}
We derive structural properties of these nuclear regions, including mass, size, rotation, and shape. By using theoretical supernova yields to model the supernova enrichment in the simulations, we obtain individual elemental abundances for Fe, O, Na, Mg, and Al. Our nuclei are systematically more metal-rich than their host galaxies, which lie on the expected mass-metallicity relation. Some nuclei have a spread in Fe \rev{and age} comparable to the \rev{massive globular clusters of the Milky Way}, lending support to the hypothesis that nuclear star clusters of dwarf galaxies could be the progenitors of these objects. None of our nuclear regions contain the light element abundance spreads that characterize globular clusters, even when a large age spread is present. Our results demonstrate that extended star formation history within clusters, with metal pollution provided solely by supernova ejecta, \rev{is capable of replicating the metallicity spreads of massive globular clusters, but still requires another polluter to produce the light element variations.}
\end{abstract}

\keywords{galaxies: formation -- galaxies: nuclei -- galaxies: star clusters: general -- globular clusters: general}

\section{Introduction}
Globular clusters have traditionally been seen as having a single-age, single-metallicity stellar population. However, recent evidence has shown that the formation process of globular clusters is more complicated \citep[e.g.][]{Gratton12}. Color-magnitude diagrams show multiple \rev{populations} of stars, as evidenced by multiple main sequences, the widening of the red giant branch, and the splitting of the horizontal branch \citep[e.g.][]{Milone17}. Additionally, spectroscopy has revealed correlations in the abundances of light elements such as N, O, Na, Mg, and Al \citep[e.g.][]{Carretta_09_data}. A large fraction of globular cluster stars are affected. 

\rev{Further deviations from a simple stellar population are present in a few of the most massive clusters, which can have spreads in both metallicity and age. The measurable spread in iron gives these objects the name ``anomalous'' or ``iron-complex'' clusters} \citep{Marino15}. They include $\omega$~Cen \citep{johnson_etal09}, M54 \citet{Carretta_etal_10_m54}, and Terzan~5 \citep{Massari_etal_10}. \rev{In clusters with an iron spread, the light element abundances are present within each peak of the metallicity distribution \citep{Marino11, Marino15}. An age spread may be present in these clusters as well. \revb{For $\omega$~Cen the age spread has several claims in the literature, from around 500 Myr \citep{tailo_etal16} to 1 Gyr \citep{Joo_Lee_13} to several Gyr \citep{Villanova_14_w_cen}.}
Terzan 5 has a much larger age spread, at around 7.5 Gyr \citep{Ferraro16}. \revb{M54 is still embedded in the stellar population of its host galaxy, making interpretation difficult, but this region has experienced star formation over at least 10 Gyr \citep{Siegel07}.} The extended formation history of these objects may imply a significantly different origin than typical globular clusters. These anomalous clusters are more similar to the nuclei of dwarf galaxies, and thus may represent transitional objects between star clusters and galaxies.}

\rev{Most nearby galaxies contain nuclear star clusters (NSCs)} \citep[e.g.][]{Leigh12}. NSCs are more prevalent in smaller galaxies ($\Mstar < 10^{10}\Msun$), but can coexist with supermassive black holes in the mass range around $10^{10}\Msun$ \rev{\citep{Ferrarese06, Seth10, Georgiev16}.} The Milky Way hosts both a NSC and a central black hole \citep[e.g.][]{FK17_orbits}. Since NSCs sit at the bottom of the galaxy's potential well, they can have inflows of gas that allow multiple star formation episodes, \rev{compared to} non-nuclear star clusters of the same mass. The Milky Way's NSC shows evidence of a complicated star formation history \citep{Do15,FK17_metallicity}. Enriched gas processed by the stars in the dense cluster is likely to remain near the cluster, allowing the later populations of stars to have enriched elemental abundances. Of particular interest are the NSCs of satellite galaxies. When a satellite galaxy merges with a larger host, its dense nucleus is likely to survive the tidal interactions. After the galaxy is stripped, the NSC may emerge as one of the host's massive globular clusters. This hypothesis is potentially capable of explaining the \rev{age and metallicity spreads} of the most massive globular clusters \citep{Freeman93,Boker08}. \rev{These spreads in metallicity may also result in spreads in light elements, potentially explaining some of the observed spreads in massive clusters.}

For this to be a viable scenario, the NSCs of the satellites need to have similar sizes, masses, and elemental abundances as the globular clusters we observe. Here we test this scenario by analyzing the results of recent ultrahigh-resolution simulations of galaxy formation by \citet{li_etal17, li_etal18}. These simulations include a novel star formation prescription, where star clusters are the unit of star formation. These star cluster particles form over time until the feedback from their stars terminates their growth. The clusters incorporate continuous enrichment of the interstellar medium and encode some intrinsic metallicity spread. This realistic modeling results in self-consistent cluster masses, \rev{age spreads, and metallicity distributions}, making the simulations an excellent tool to study this problem.

In this paper we identify nuclear regions in the simulations, then compare their masses and sizes to a sample of observed NSCs \citep{Georgiev16}. We check the mass-metallicity relation for our galaxies and nuclear regions against the observed relation for dwarf galaxies \citep{Kirby13}. We use published supernovae yields \citep{Nomoto06} to make predictions of the individual elemental abundances for the stars in the nuclear regions. 
\rev{We then examine the age and metallicity spread of the nuclear regions, as a direct comparison to the observed massive globular clusters. Lastly, we compare the spreads in light elements to the observed spreads found in globular clusters}. 

\section{Structure of Nuclear Clusters in Simulations}

The suite of cosmological simulations by \citet{li_etal17, li_etal18} was run using the Adaptive Refinement Tree (ART) code \citep{kravtsov_etal97,kravtsov99,kravtsov03,rudd_etal08} in a 4 Mpc comoving box, with initial conditions selected to produce one Milky Way-sized galaxy at $z=0$ along with several satellite galaxies. The ART code solves equations for the gravitational dynamics of the stars, dark matter, and gas, as well as the hydrodynamics of the gas component. The code utilizes Adaptive Mesh Refinement (AMR) to reach very high spatial resolution. The current suite has a maximum resolution of $L_{\rm cell} = 3-6$ physical pc (not comoving), which is high enough to resolve giant molecular clouds. The code calculates run-time transport of UV radiation \citep{ngnedin_abel01} from both local stellar sources \citep{n_gnedin_14_reionization} and the extragalactic background \citep{haardt_madau_01}. A non-equilibrium chemical network is used to model the various ionization states of hydrogen and helium, and the formation and destruction of molecular hydrogen \citep{ngnedin_kravtsov11}. Stellar feedback comes from the energy and momentum from supernovae, as well as winds and radiation pressure from young massive stars, which is calibrated by stellar population synthesis models. Chemical enrichment includes contributions from supernovae types II and Ia. The most novel aspect of these simulations is the explicit modeling of stars forming in clusters with a spectrum of masses that matches observations of young star clusters in the nearby universe. Growth of star clusters is terminated by their own feedback, and cluster masses are thus calculated self-consistently. This current suite has completed runs to redshift $z\approx 1.5$.

We use the outputs of several runs with different value of the local star formation efficiency per free-fall time, $\epsff$. All runs start from the same initial conditions and have the same physics, including the supernova momentum boost factor $f_{\rm boost}=5$. This boost factor accounts for the enhanced momentum feedback of clustered SNe vs isolated SNe, and also compensates for the momentum loss due to advection errors as the SN shell moves across the simulation grid \citep[for details see][]{li_etal18}. We take four runs with a constant value of $\epsff$ randing from 10\% to 200\% (SFE10, SFE50, SFE100, and SFE200), plus one run with turbulence-dependent $\epsff$ (SFEturb, with the median of about 3\%). One additional run, SFE50-3SNR with $f_{\rm boost}=3$, was chosen to test the sensitivity of results to the strength of feedback. Despite the different choices of $\epsff$, all runs with $f_{\rm boost}=5$ reproduce the galaxy stellar mass and star formation rate as expected from the abundance matching technique; the SFE50-3SNR run has an elevated star formation rate because of weaker feedback. While $\epsff$ has little effect on the global galaxy properties, the properties of individual star cluster particles (such as the cluster mass function, maximum cluster mass, and the cluster formation timescale) depend strongly on this parameter. For this reason, we plot each of these runs separately, although it will be apparent that our results are not affected by the numerical differences between the runs. 

\subsection{NSC Masses and Sizes}
\label{ssec:mass_size}

We use the ROCKSTAR halo finder \citep{rockstar} to identify the dark matter halos in the simulation outputs. Then we find the stellar center of the galaxy by doing an iterative centering process. To account for discreteness due to the small number of stellar particles in the center, we use kernel density estimation (KDE) where we smooth each star particle with a 3D Gaussian kernel, then sum the contributions to create a full stellar density field. We start by smoothing the star particles with a large kernel ($\sigma=2\,$kpc). We pick the location with the highest density, then recalculate the stellar density around that point using a kernel that is 3 times smaller. We repeat until we reach the resolution limit of the simulation, given by the cell size at the finest refinement level. The smallest kernel uses $\sigma = L_{\rm cell, min} =3\,$pc. This iterative process has the effect of first picking out large-scale galactic structure, then focusing in on the locations with the highest stellar density within those larger scale dense regions. The location of the absolute highest stellar density sometimes is in a very massive young cluster in the outskirts of the galaxy, but the algorithm avoids these clusters in favor of true nuclear clusters. Using the large scale galactic structure is essential to this process.

Next we define the plane of the galaxy. Following the formalism in \citet{Zemp11}, we use the inertia matrix to calculate the axis ratios of the galaxy. The inertia tensor can be written as
\begin{equation}
	I_{ij} = \frac{\sum_k M_k \, \mathbf{r}_{k,i} \, \mathbf{r}_{k,j}}{\sum_k M_k}
    \label{eq:inertia_matrix}
\end{equation}
where $M_k$ is the mass of the $k$-th particle, and $\mathbf{r}_{k,i}$ is the $i$ component of the position of the $k$-th particle. We use only the star particles to calculate the axis ratios, as we are interested in the stellar component. The eigenvalues of this tensor are $a^2/3$, $b^2/3$, and $c^2/3$, where $a$, $b$, and $c$ are the semi-principal axes, and $a \geq b \geq c$. We define the normal vector to the plane of the galaxy to be the eigenvector corresponding to the smallest eigenvalue ($c^2/3$). 

We project all the star particles onto the plane of the galaxy, then perform a two dimensional KDE to obtain the surface density profile. For star particles within the inner 12 pc, we take the Gaussian kernel with a one sided width of 3 pc ($L_{\rm cell,min}$). Outside of 12 pc, we use a larger 6~pc kernel ($L_{\rm cell,max}$) to reduce counting noise. The surface density is calculated by integrating the KDE density over an annulus to get the mass, then dividing by the area of the annulus. This KDE estimate is used only for the central 100 parsecs, while outside this radius we simply bin the star particles. 

\begin{figure}
	\includegraphics[width=0.48\textwidth]{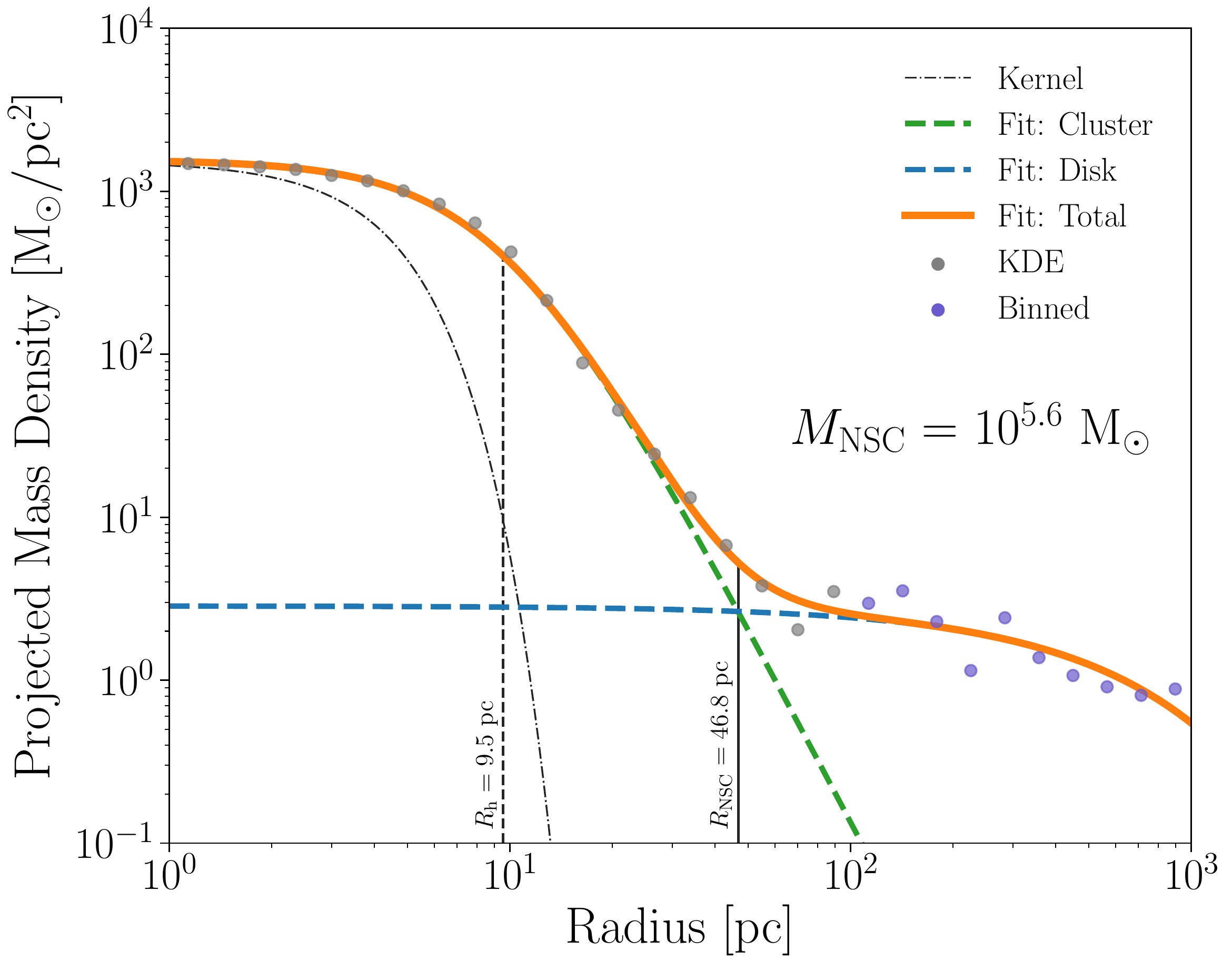}
    \caption{An illustration of a two-component fit of the surface density profile of a typical galaxy. The two components of the fit are a nuclear region and an outside disk. Vertical lines show the half-mass radius $\Rh$ and the full extent of the nuclear cluster $\Rnsc$. The cluster mass is indicated at the bottom right.}
    \label{fig:profile}
\end{figure}

To find the nuclear cluster, we perform a two-component fit of the stellar density profile. Galaxies at these redshifts are expected to be mostly disky with an additional central component (nucleus or bulge), so we use the sum of an exponential disk and a Plummer sphere:
$$ \Sigma_{\rm disk}(R) \propto \exp{(-R/a_d)}, \quad 
   \Sigma_{\rm cluster}(R) \propto [1+(R/a_c)^2]^{-2}, 
$$
where $a_d$ and $a_c$ are the scale radii for the two components \citep{Dejonghe87}. We define the size of the nuclear region, $\Rnsc$, to be the radius where $\Sigma_{\rm cluster}(\Rnsc) = \Sigma_{\rm disk}(\Rnsc)$. \autoref{fig:profile} shows an example of this fitting process.

To estimate the uncertainty of our determination of $\Rnsc$ after marginalizing over the other structural parameters, we construct a multivariate Gaussian distribution with dimensions given by the covariance of the parameters of the decomposition (normalization and scale radius of both components). We then sample from this multivariate Gaussian and re-calculate the radius of the nuclear region for each realization. We take the range that encloses 68.3\% of these radii to be the error of $\Rnsc$.

The mass of the nucleus then is the sum of the star particles within this spherical region, $\Mnsc\equiv \sum_k M_k(r_k<\Rnsc)$. We do not subtract off the disk component within this region. We also calculate the half-mass radius of the nuclear cluster, defined as
\begin{equation}
  M(\Rh) \equiv \frac{1}{2} M(\Rnsc), \quad
  M(R) = \int_0^{R} \Sigma_{\rm KDE}(R) \, 2\pi R\, dR
  \label{eq:rh}
\end{equation}
where $\Sigma_{\rm KDE}(R)$ is the KDE surface mass density as described above. This half-mass radius is calculated in projection, unlike the mass, for a more direct comparison with observations. The errors on the mass and half-mass radius are estimated by perturbing $\Rnsc$ according to its error.

\begin{figure}
	\includegraphics[width=0.48\textwidth]{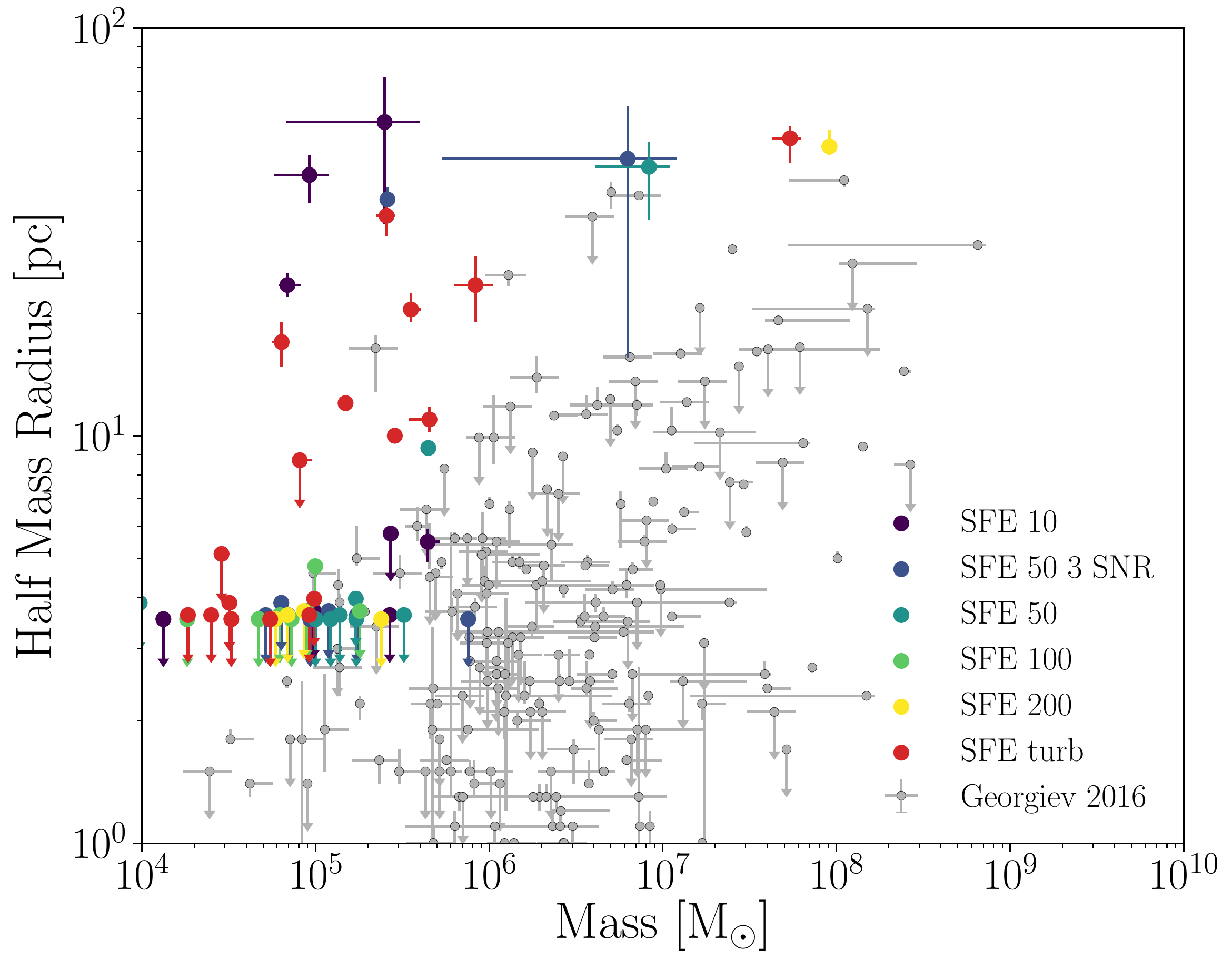}
    \caption{Masses vs. half-mass radii of the nuclear regions of our simulated galaxies (colored symbols). Nuclear regions where one star particle contributes more than half of the total mass are shown as having upper limits on $\Rh$. Additionally, an upper limit is given to one nuclear region with a small density contrast, where the fit parameters were uncertain enough that the 1$\sigma$ limits on the mass and radius included zero. As a comparison, the sizes and masses of nuclear star clusters in late type galaxies presented in \citet{Georgiev16} are shown in grey.}
    \label{fig:mass_size}
\end{figure}

\autoref{fig:mass_size} shows the masses and sizes of the galactic nuclei in the simulations, along with observed NSCs in nearby late-type galaxies of intermediate and low mass ($\Mstar < 10^{11}$) from \citet{Georgiev16}. We excluded one nuclear region with mass above $10^9\Msun$, in the SFE50-3SNR run that had insufficiently strong feedback to suppress the formation of a very dense bulge. Simulated nuclear regions with half-mass radii near 3~pc are due to a single star particle dominating the central mass distribution. For these regions, where a single particle contributes more than half of $\Mnsc$, we take the $\Rh$ given by \autoref{eq:rh} to be an upper limit because they just reflect the size of the smoothing kernel. At low masses, there are many of these galaxies where a single star particle dominates. These unresolved clusters are consistent with many of the compact clusters in the \citet{Georgiev16} sample. 

However, the nuclei that are resolved in the simulations tend to have $\Rh$ larger than observed clusters of the same mass. This overestimate is particularly noticeable at low masses, and can be an order of magnitude or more. At high masses the discrepancy is a factor of several.

\rev{As discussed above, we select nuclear regions by means of a two-component fitting procedure. While this procedure generally provided sizes larger than the observed radii of the nearby NSCs, any more complicated model fit would be unwarranted due to the irregular structure of our galaxies. Our fit also appears to be robust and nuclear regions clearly defined. The central NSC density exceeds the extrapolated central density of the disk component typically by a factor of 10 or greater. }

\rev{However, the physical interpretation of these components is not always clear. Strong feedback gives our galaxies an irregular structure at the redshift $z\approx 1.5$ of our last output, making the identification of a center unclear in some cases. Some galaxies had a roughly constant density in the center, with one massive star particle that was chosen by our algorithm to be the center. Others had a more extended central component without a strongly peaked center, where the middle was chosen to be the center. Even in these situations, however, a central component is present and can represent a nuclear region, even if it is too large to be a typical NSC. Visually inspecting the profiles indicated that the returned sizes were an acceptable description of this central component. These results indicate that higher resolution or improved stellar physics are required to properly model NSCs.}

\rev{Interestingly, we find that although NSC sizes are generally overestimated in the simulations, the overall galaxy sizes match relatively well with the observed sample of \citet{VanDerWel14}. For late-type galaxies with $9.0 < \log \Mstar < 9.5$ at redshift $1.5 < z < 2$, the median effective radius is 2.1 kpc, compared to 2.8 kpc for our simulated galaxies in the same range of mass and redshift. The distribution of sizes in the observed sample (68\% lie between 1.2 and 3.7 kpc) is broader than in the simulations, which range from 2.5 to 3.3 kpc. The sizes of our simulated galaxies are thus statistically consistent with the available observations.}

\subsection{Ellipticity and Rotation}

With the nuclear region defined, we can assess the spatial and kinematic structure of the collection of stars in this region. Using the inertia tensor (\autoref{eq:inertia_matrix}), we calculate the axis ratios of the nuclear regions.

\begin{figure}
	\includegraphics[width=0.48\textwidth]{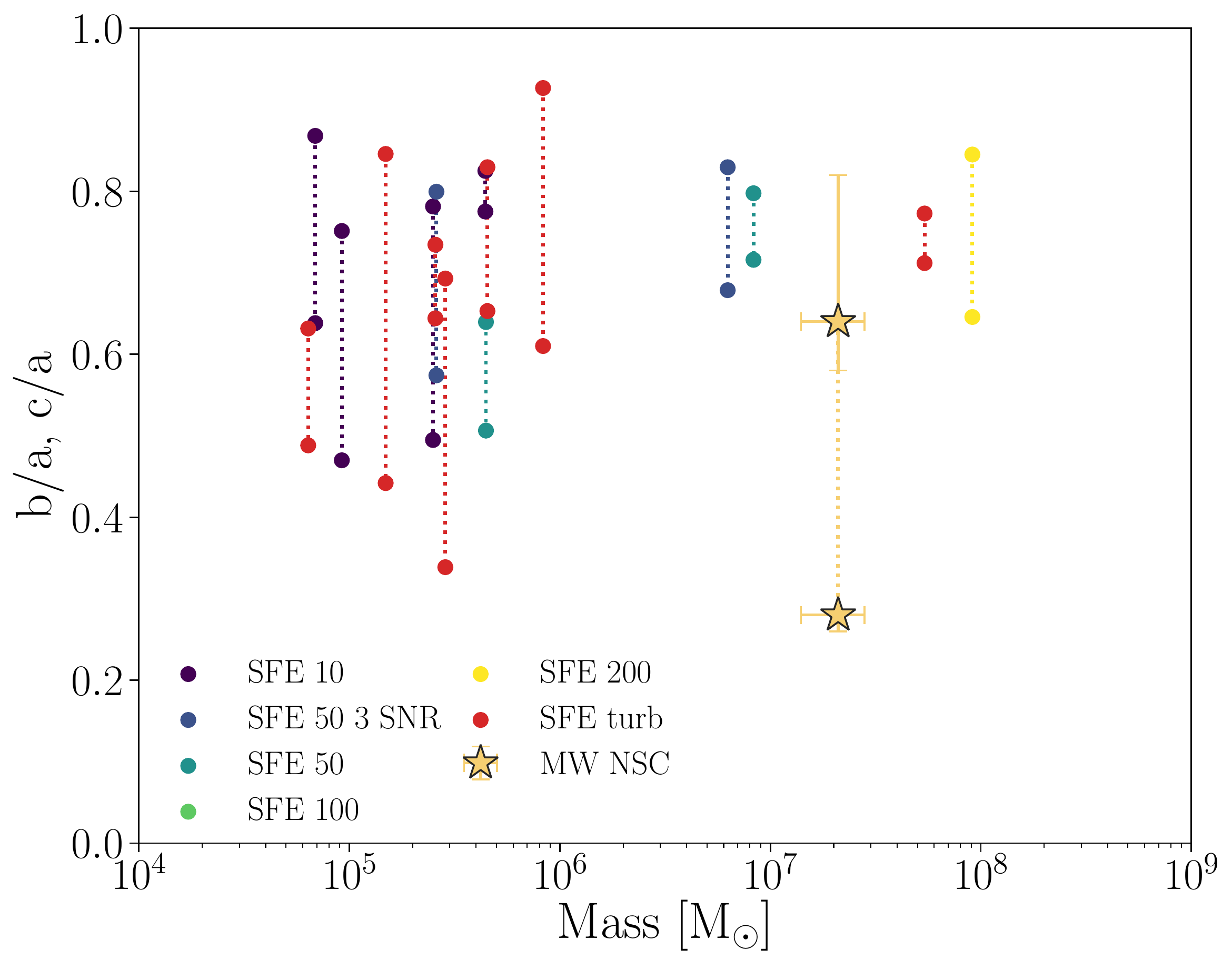}
    \caption{Axis ratios of our simulated galaxies as a function of mass of the nuclear region. Each galaxy has two points connected with a line. The upper shows $b/a$, while the lower shows $c/a$. Nuclear regions that had too few star particles to properly calculate the axis ratios are not shown. The Milky Way NSC is shown as a star, for comparison \citep{FK17_orbits}.}
    \label{fig:axis_ratios_vs_mass}
\end{figure}

\autoref{fig:axis_ratios_vs_mass} shows the axis ratios as a function of the NSC mass. Nuclear regions that have less than 10 star particles are not shown, as they have too few particles to reliably determine the axis ratios. Our nuclear regions are mildly triaxial, but all of them are less flattened than the Milky Way NSC \citep{FK17_orbits}. 

To investigate whether the flattening is caused by rotation, we calculate the rotation velocity and the residual velocity dispersion of the NSCs. We define the plane of the galaxy in the same way we did above, based on the inertia tensor. We found that this definition works better than using the angular momentum vector of the nuclear regions, because the latter is relatively small and does not determine the structure of the NSCs. We find that the direction of the nuclear angular momentum is misaligned with the normal to the plane of the galaxy by between 20 and 90 degrees, and has no dependence on cluster mass.

\begin{figure}
	\includegraphics[width=0.48\textwidth]{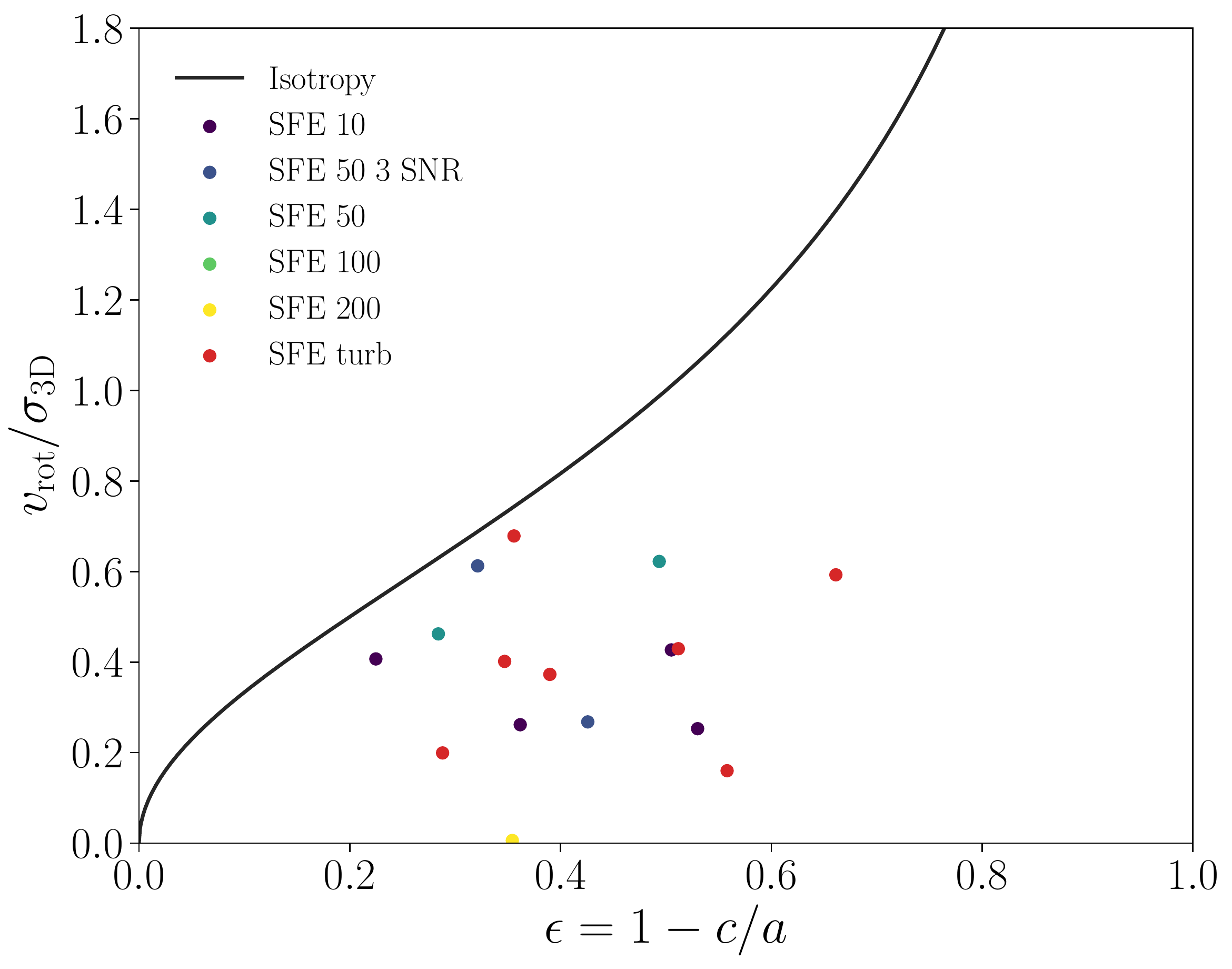}
    \caption{The amount of rotational support of the nuclear regions as a function of their ellipticity. The black line shows the expected values for a system with an isotropic velocity dispersion that is flattened only by rotation. Most points lie below this line, indicating that the nuclear regions are intrinsically non-spherical.}
    \label{fig:axis_ratios_velocity}
\end{figure}

The rotation velocity is calculated as the mass-weighted tangential velocity component of the star particles within the nuclear region. \autoref{fig:axis_ratios_velocity} shows the ratio of rotational velocity to 3D velocity dispersion as a function of ellipticity, defined as $\epsilon \equiv 1-c/a$. The black line shows the expected values for a system with an isotropic velocity dispersion that is flattened only by rotation. This relation can be approximated as
\begin{equation}
  \frac{v_{\mathrm{rot}}}{\sigma_{3D}} \approx \sqrt{\frac{\epsilon}{1-\epsilon}}
  \nonumber
\end{equation}
\citep[e.g.,][]{mo_vandenbosch_white_book}.
The simulated nuclei are predominately below this line, indicating that their non-sphericity is an intrinsic shape, not due to rotation. The accretion of particles composing the NSCs must have proceeded anisotropically. Indeed, we find that for nearly all clusters the radial velocity dispersion is larger than the tangential dispersion by up to a factor of two.

\subsection{Metallicity}

As described above, our suite of simulations tracks enrichment from Type Ia and II supernovae, but not AGB winds. We calculate the mass-weighted total metallicity as the sum of all metals divided by the mass of all star particles in the nuclear regions. 

To calculate the spread in metallicity, we consider two sources of variance: the internal spread within a star cluster particle due to self-enrichment (recorded in the simulation runtime as the cluster is forming), and the dispersion of final metallicity among star particles.

To calculate the internal spread $\sigma_{Z,k}^2$ within the $k$-th particle, we transform the spread in overall metallicity from SNII. The details of how this is derived from the simulation outputs is described in H. Li \& O. Gnedin (2018, in preparation), but we summarize the key points here. The star particles in the simulation contain a variable $M^{ZZ} = \sum m_i Z_{i, {\rm II}}^2$ that can be used to calculate the metallicity variance of a single star particle without needing its full accretion history. Here we are summing over all accreted mass elements $m_i$ of a specific cluster particle as it is forming at time steps $i$. The final particle mass is $M_k = \sum m_i$. We only consider the spread in metallicity from Type II supernovae, since the delay time for SNIa is significantly longer than the formation timescale of star clusters and therefore enrichment from Type Ia supernovae will not be relevant for this spread. The variance of the metallicity of a single star particle $k$ is defined as:
\begin{equation}
	\sigma_{Z,k}^2 = \frac{\sum_i m_i Z_{i, {\rm II}}^2}{\sum_i m_i} - \left(\frac{\sum_i m_i Z_{i,{\rm II}}}{\sum_i m_i}\right)^2
    = \frac{M_k^{ZZ}}{M_k} - Z_{k,\rm II}^2.
\end{equation}

We then combine this with the dispersion in metallicity among cluster particles to obtain the total metallicity spread:
\begin{equation}
  \Delta Z^2 = \frac{\sum M_k \, \sigma_{Z,k}^2}{\sum M_k} + \frac{\sum M_k \left(Z_k - \overline{Z}\right)^2}{\sum M_k}
  \label{eq:met_spread}
\end{equation}
where $\overline{Z}$ is the mass-weighted average metallicity of the nuclear region:
\begin{equation}
  \overline{Z} \equiv \frac{\sum M_k Z_k}{\sum M_k}.
\end{equation}

\begin{figure}
	\includegraphics[width=0.48\textwidth]{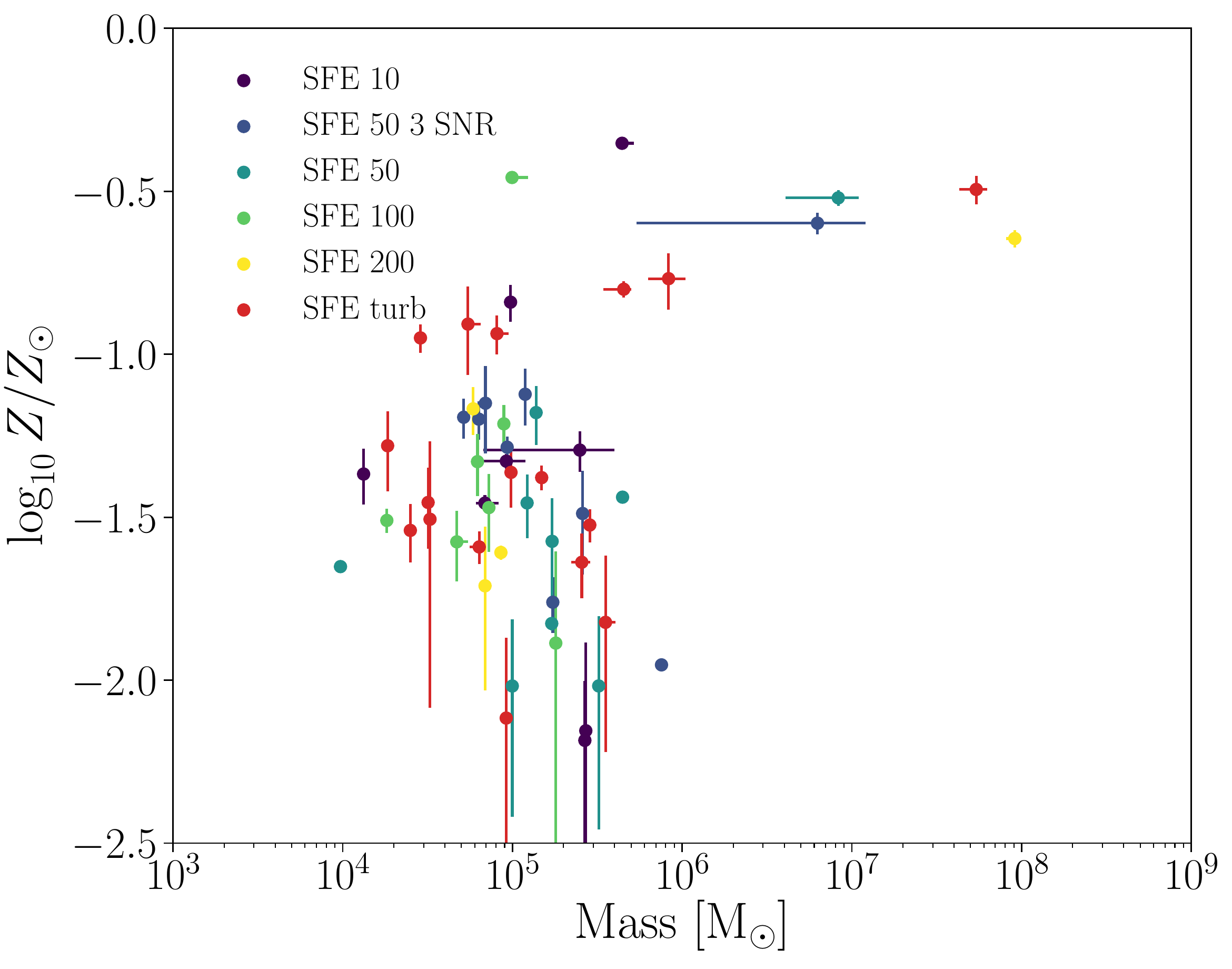}
    \caption{Total metallicity of the nuclear regions as a function of their mass. Errorbars cover the interval $\log(\overline{Z}-\Delta Z)$ to $\log(\overline{Z}+\Delta Z)$.
    }
    \label{fig:mass_z}
\end{figure}

\autoref{fig:mass_z} shows the metallicities of the nuclear regions plotted against their mass. We represent the metallicity spread with errorbars that cover the interval $\log(\overline{Z}-\Delta Z)$ to $\log(\overline{Z}+\Delta Z)$. Our nuclear regions have a wide spread of metallicities in the range $-2 < \feh < -0.25$. Other then the most massive nuclear regions typically having higher metallicity, there is no correlation between metallicity and cluster mass.

\section{Elements}

\subsection{Yields}

To turn the total metallicity into abundances for individual elements, we use computed supernova yields in the literature. We take the SNII yields of \citet{Nomoto06} and integrate them over the \citet{Kroupa01} IMF from 10 to 40 $\Msun$, using a 50\% hypernova fraction for stellar masses where hypernova yields are available. Since the \citet{Nomoto06} yields are provided at several fixed progenitor metallicities $Z=$ 0, 0.001, 0.004, and 0.02, we interpolate between these models to get the yield at any metallicity. We use the W7 model of \citet{Iwamoto99} for SNIa yields at all metallicities.

To get the elemental abundances of a given star particle, we use the SN yields and scale them to produce the appropriate mass of metals, as set by the $Z_{\rm Ia}$ and $Z_{\rm II}$ variables in the simulation. 

The two sets of published SN yields depend on the metallicity of the supernova progenitor star. Since this information is not recorded in the simulation output, we are using the metallicity of the star particle itself as the metallicity of the progenitor. This may overestimate the progenitor metallicity. However, as we discuss later, we use elements whose yields do not vary strongly with metallicity of the progenitor, reducing the effect on our results.

\begin{figure*}
	\centering
	\includegraphics[width=0.74\textwidth]{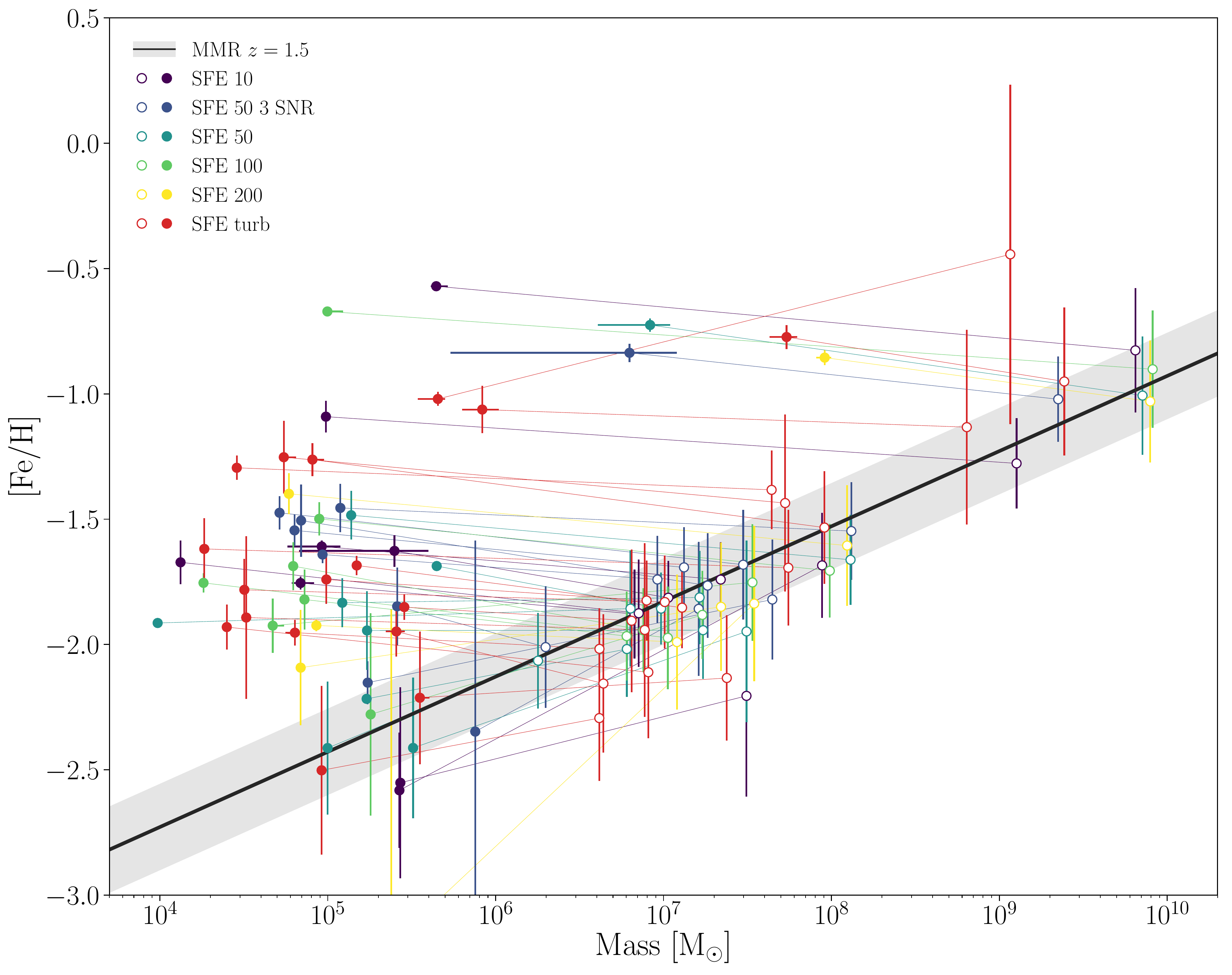}
    \caption{Mass-metallicity relation (MMR) for the whole simulated galaxies (open circles) and their nuclear regions (filled circles). The lines connect the nuclei to their host galaxies. Errorbars in [Fe/H] are the spread in [Fe/H] in a nuclear region or galaxy (\autoref{eq:elt_disp}). The black line and shaded region show the MMR for nearby dwarf galaxies with its associated RMS scatter of 0.17~dex, shifted down by 0.36~dex to account for evolution in the MMR to $z=1.5$.}
    \label{fig:mass_metallicity}
\end{figure*}

\subsection{Mass - Metallicity Relation}

By using these yields, we calculate a mass-weighted value of $\feh$ for both the star particles in the nuclear regions and the stars in the whole galaxy. The average contribution from all star cluster particles is
\begin{equation}
  \overline{\feh} = \log \left[ \frac{\sum_k M_k (Z_{k,Ia} f_{\rm Fe,Ia} + Z_{k,II} f_{\rm Fe,II})}{\sum_k M_k\, X_k} \right] - \log \left[\frac{Z_\odot f_{{\rm Fe}\odot}}{X_\odot} \right]
\end{equation}
where $f_{\rm Fe}$ is the mass fraction of the metals that is Fe. The numerator in each fraction is the mass of iron, while the denominator is the mass of hydrogen. For the Sun we use the solar abundances of \citet{Grevesse_Sauval_98}, which are $Z_\odot = 0.0169$ and $f_{\rm{Fe}} = 0.0757$.

We then transform the metallicity spread (\autoref{eq:met_spread}) into spread in $\feh$ by using:
\begin{equation}
  \Delta\feh^2 = \Delta Z^2 \left( \frac{d\feh}{d Z_{\rm II}} \right)^2.
\label{eq:elt_disp}
\end{equation}
We calculate this derivative numerically using interpolation of the published SN yields. While this equation is written using $\feh$, it will also be used to calculate spreads in other elemental abundances.

We calculate the mean and total variance in $\feh$ for both the nuclear regions and the galaxy as a whole. The results of these calculations are shown in \autoref{fig:mass_metallicity}, along with an empirical mass-metallicity relation (MMR) for dwarf galaxies from \citet{Kirby13}, shifted down by 0.36~dex to account for evolution in the MMR from $z=0$ to $z=1.5$, \rev{based on the interpolation of the results of \citet{Mannucci09} for Lyman-break galaxies at $z\approx 3$}.

The model galaxies lie near the expected $z=1.5$ relation, without any fitting, which provides evidence that our simulations are modeling the chemical enrichment of galaxies correctly. The nuclei are systematically more metal rich. This indicates that our galaxies have a metallicity gradient, consistent with observations \citep[e.g.][]{Moustakas_etal_10}. The predicted iron abundances of NSCs span a wide range $-2.5 < \feh < -0.5$, uncorrelated with their mass.

\subsection{Effects of Extended Star Formation}

To examine the effect of an extended star formation history, we first determine the time it took to form the bulk of the stellar mass in each nuclear region. To minimize the effect of few particles born much sooner or later than the bulk of the particles, we pick the smallest time interval in which 90\% of the mass was formed. This interval is calculated using only the birth time of the particles, not including the time each star particle took to form. To account for this additional time, we add the maximum duration of star particle formation within the cluster ($2 {\rm \ Myr} \lesssim \tau_{\rm dur} \lesssim 10 {\rm \ Myr}$; \citealt{li_etal17}) to the initial time interval. 

\begin{figure}
	\includegraphics[width=0.48\textwidth]{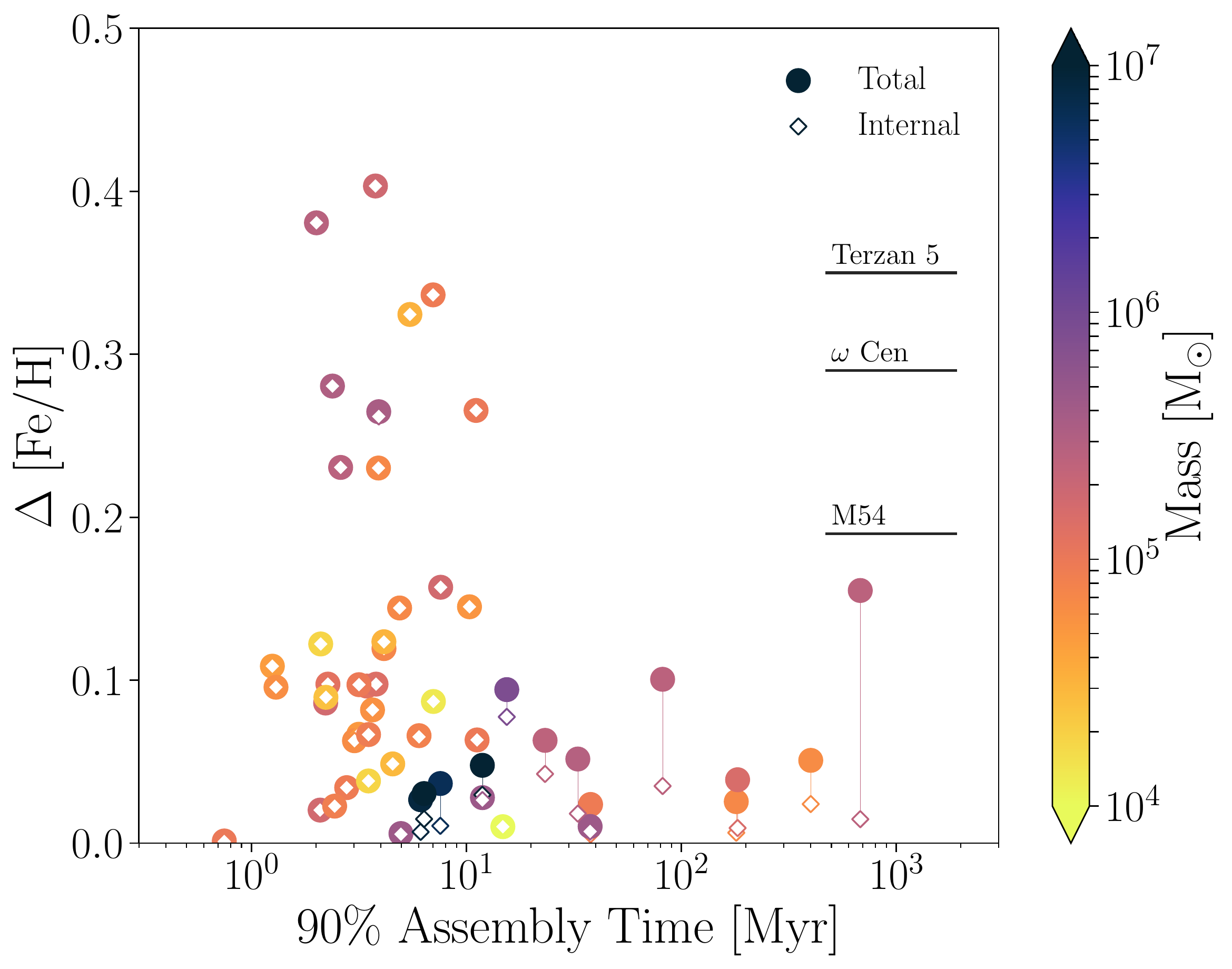}
    \caption{Dispersion of Fe in the nuclear regions as a function of the time required to assemble 90\% of their mass. The contribution from internal star particle metallicity spread is shown by hollow diamonds and labeled ``Internal'', while the total dispersion including spread among the multiple star particles in the nuclear region is shown by filled circles and labeled ``Total''. The internal and total components for each NSC are connected by a vertical line. The points are color-coded by the mass of the nuclear region. The horizontal lines on the right show the observed Fe spreads in massive globular clusters of the Milky Way (see text for references).}
    \label{fig:sfh_time}
\end{figure}

\autoref{fig:sfh_time} shows that the iron spread in the nuclear regions does not correlate with the length of star formation history. In fact, several NSCs with the longest assembly time have relatively small spread, $\Delta\feh < 0.15$, most of which is due to different metallicities of its constituent particles, rather than the intrinsic spread within each particle. The largest spread is seen in NSCs that took only 2-3 Myr to form, where all the spread is internal. These regions tend to be low metallicity ($\feh < -1.8$) and consist of only one or few star particles. 

The largest predicted iron spread is comparable with the spread in three massive globular clusters in the Milky Way where it is clearly detected: $\omega$~Cen, M54, and Terzan~5. \revb{To make a fair comparison to our models,} we \revb{define} the observed spread as the RMS scatter of the $\feh$ content of the stars in each cluster. \revb{This value is calculated for M54 by \citet{Carretta_etal_10_m54} and for Terzan 5 by \citet{Massari_etal_10}, while we calculate it for $\omega$~Cen using the data from \citet{Marino11}}
\rev{In \autoref{fig:sfh_time} we do not show the observed age spreads, only the metallicity spreads. The age spreads are uncertain, with inconsistencies between different studies of $\omega$~Cen \citep{Joo_Lee_13, Villanova_14_w_cen,tailo_etal16}. In our simulations we are able to detect both age and metallicity spreads significantly below the observational limits, which remain relatively large.} \revb{However, the simulated assembly timescales are within the inferred age spread of stellar populations in $\omega$~Cen of $\sim 500$~Myr \citep{tailo_etal16}, while being shorter than the age spreads in M54 and Terzan 5 \citep{Siegel07,Ferraro16}}. \revb{As our simulations only reach $z \approx 1.5$ ($t \approx 4.3$ Gyr),  we are not able to model the large age spreads present in these two clusters.} 
\revb{These results} lend support to the hypothesis that stripped NSCs could become progenitors of the anomalous globular clusters. There are still quantitative differences that need to be explored in future work: the nuclei with largest iron spread all have lower metallicity than the three observed anomalous clusters and lie in the lower mass range below $10^6\Msun$.  

\subsection{Reliability of Light Element Yields}

When determining which elements can be modeled reliably by our simulations, we consider several aspects. 

(i) Our simulations only model supernovae, not AGB stars or other sources of chemical enrichment, so we select elements without strong AGB contributions.

(ii) As discussed above, we are using the metallicity of the star particle itself as the metallicity of the supernova progenitor star. To reduce the impact on our results, we prefer elements whose yields do not vary strongly with metallicity of the progenitor.

(iii) Since the two sets of yields give generally different results, we prefer the elements with low spread between the different model sets.

(iv) We select the elements whose abundances match observations, whether in direct observation of SN ejecta, observations of low metallicity stars, or chemical evolution models.

\begin{figure}
	\includegraphics[width=0.48 \textwidth]{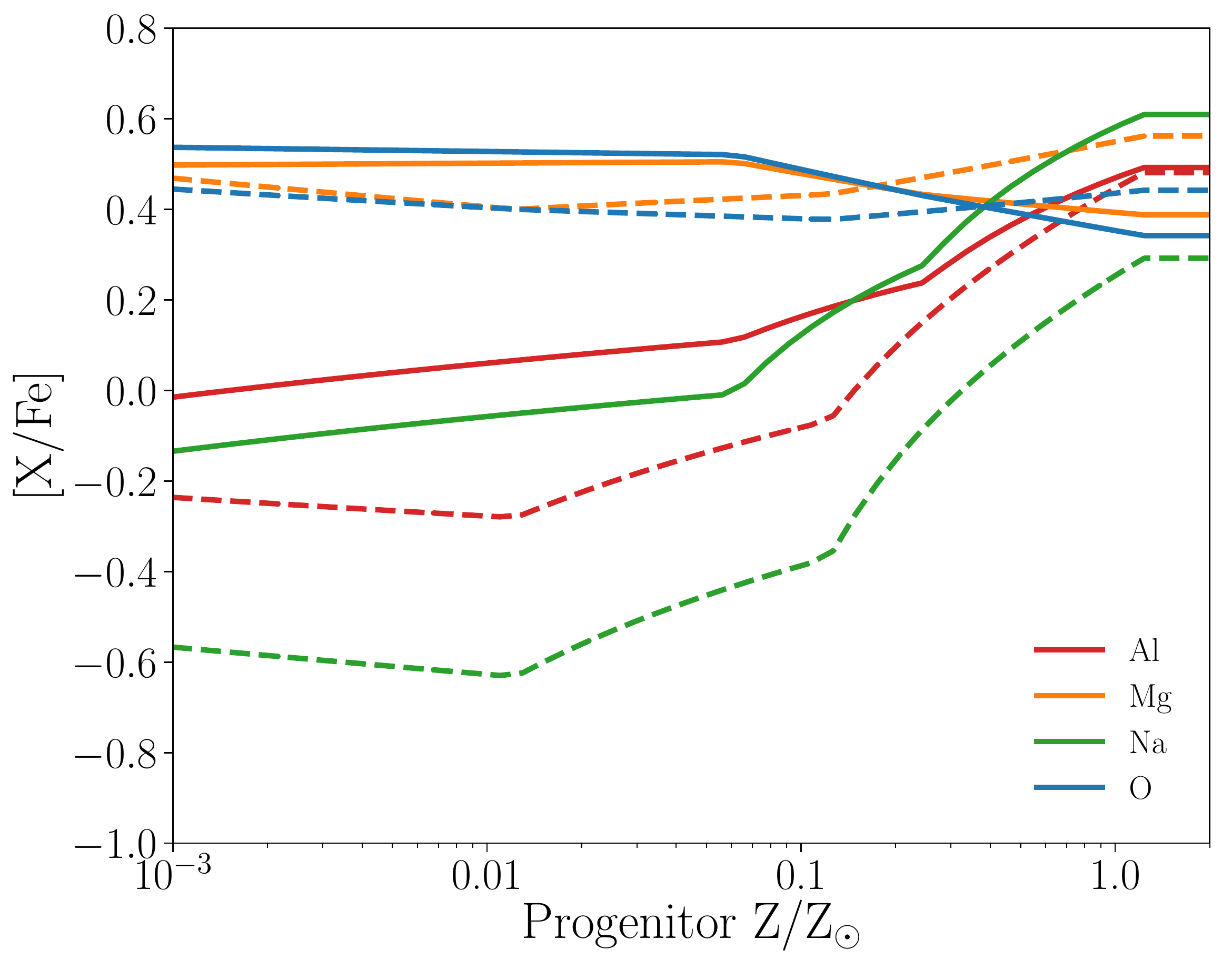}
    \caption{Abundance ratios of the ejecta of SN type II as a function of metallicity of the SN progenitor, for two yield models: solid lines are for \citet{Nomoto06}, dashed lines are for \citet{WW95}. Note that we have adjusted the WW Fe yields down by a factor of 2 and the Mg yields up by a factor of 2, as suggested by other authors \citep{Timmes95,Wiersma09,Andrews17}. We aim to study elements that do not vary strongly with $Z$ and have a low discrepancy between the models.}
    \vspace{1mm}
    \label{fig:sn_z}
\end{figure}

\autoref{fig:sn_z} shows how our two yield sets \citep{WW95,Nomoto06} vary with metallicity, to address the second and third points. \autoref{tab:gc_relevance} describes in more detail several elements that are relevant for globular clusters and NSCs, the production source of those elements, and comments on the reliability of the yields. This table includes only elements that we deemed relevant for our analysis. We do not discuss elements such as C and N, which have significant AGB contributions, or other elements that are not strongly involved in the phenomenon of multiple populations. 

To evaluate the reliability of our use of supernovae yields to extract elemental abundances for these light elements, we surveyed the literature for studies examining this issue. \citet{Romano10} and \citet{Andrews17} put different sets of yields through the same chemical evolution model, then compared the result to observations, showing how different model choices affect the resulting abundances. These papers show both the scatter among different models and the differences between models and data, which are important for assessing reliability of the yields. \citet{Andrews17} and \citet{PignatariNuGrid} describe the production sites of the various elements. \citet{Wiersma09} implement the yields into cosmological smoothed-particle-hydrodynamics simulations and show in their figure A4 the variations caused by different yields. 

The synopsis of these studies is that Oxygen is modeled reliably. It matches observations, has a weak metallicity dependence, and has a reasonably small spread between different yield sets. Magnesium is similar to Oxygen. Aluminum and Sodium have much stronger metallicity dependences and wider spreads, making their predicted abundances less certain. 

\subsection{Predicted Correlations of Element Abundances}
\label{ssec:correlations}

\begin{figure}
	\includegraphics[width=0.48\textwidth]{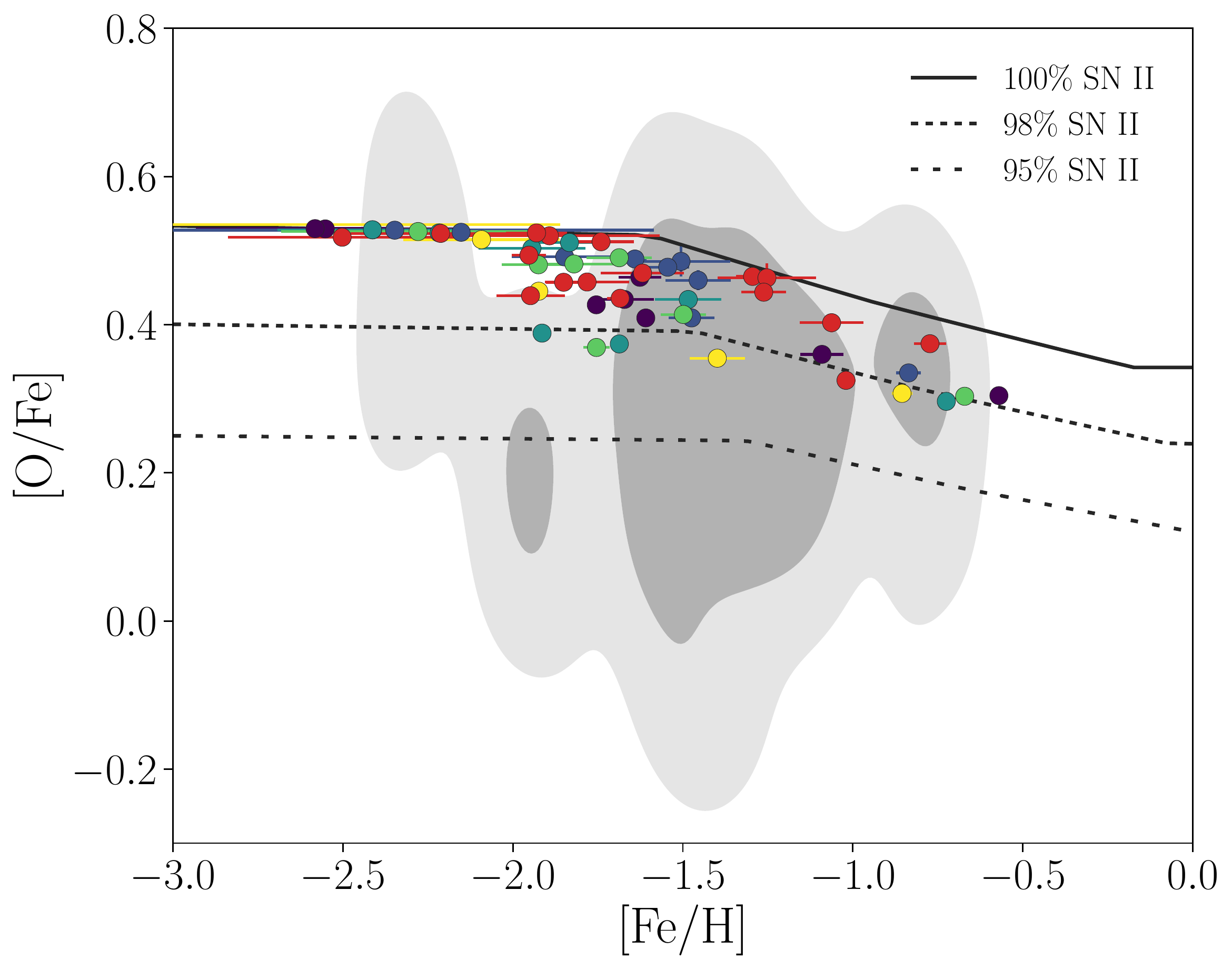}
    \caption{Oxygen abundances for model nuclear regions. The dark (light) contours show the location of 50\% (90\%) of globular cluster stars from \citet{Carretta_09_data}. Black lines show the elemental ratios of the SN yields, for different percentage contributions of SN type II.}
    \label{fig:o_fe}
\end{figure}

\begin{figure}
	\includegraphics[width=0.48\textwidth]{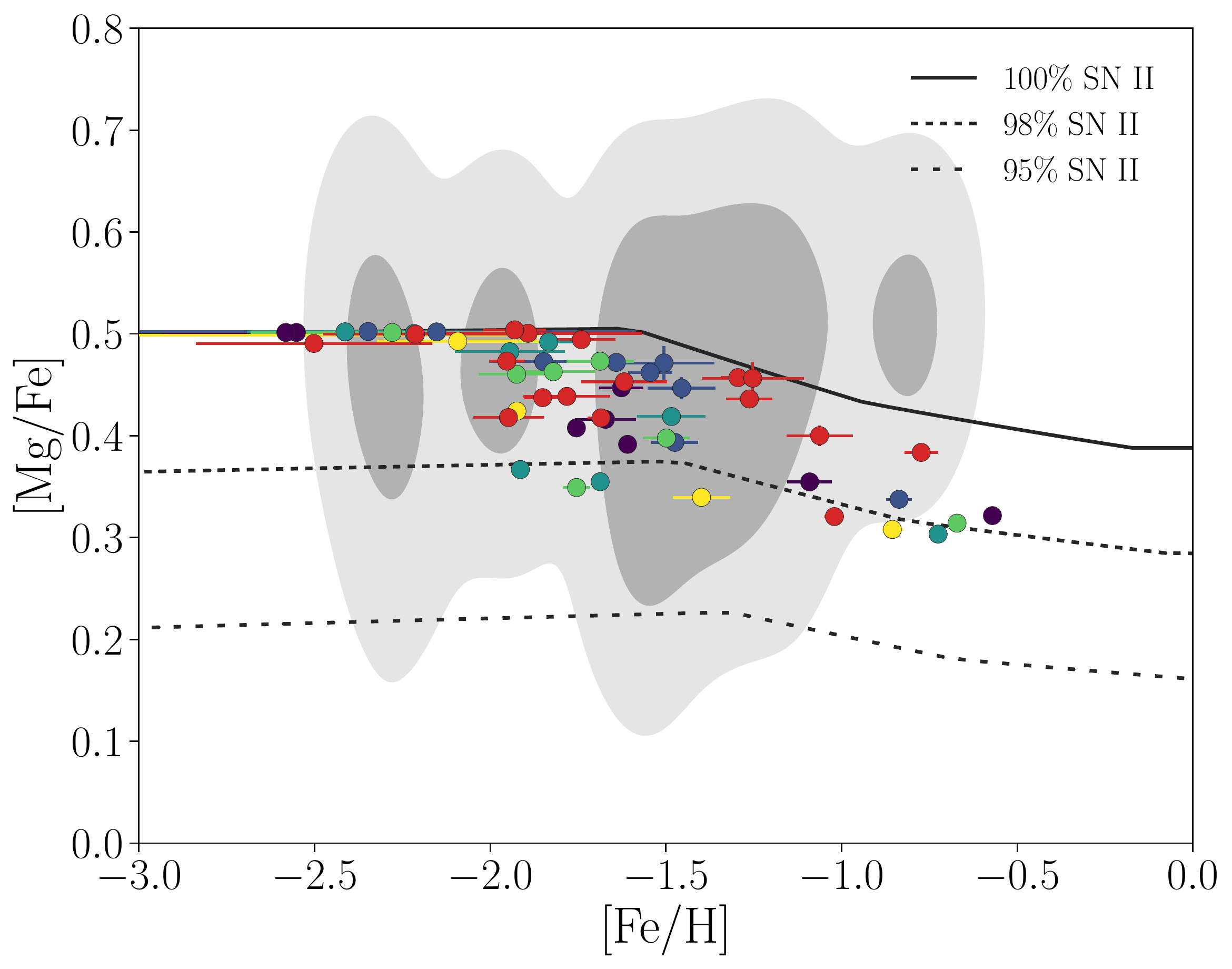}
    \caption{Same as \autoref{fig:o_fe}, but for Magnesium.}
    \label{fig:mg_fe}
\end{figure}

To test whether our nuclear regions may have properties similar to the observed multiple populations in globular clusters, we examine their detailed elemental abundances. First we compare the overall normalization of the abundances as a function of metallicity. \autoref{fig:o_fe} and \autoref{fig:mg_fe} show [O/Fe] and [Mg/Fe] for our nuclear regions and compare them to the compilation of 202 red giants from 17 globular clusters presented in \citet{Carretta_09_data}. Because of the limited sample size of the available measurements, we combine the available data from all clusters and plot the regions enclosing 50\% and 90\% of all stars. We compare this total range of observed abundances with our model predictions.

In each plot a solid line shows the track the objects would follow if their metallicity came entirely from Type II supernovae. Points below this line indicate some enrichment from Type Ia supernovae, which contribute Fe, but not O or Mg. 

\begin{figure}
	\includegraphics[width=0.48\textwidth]{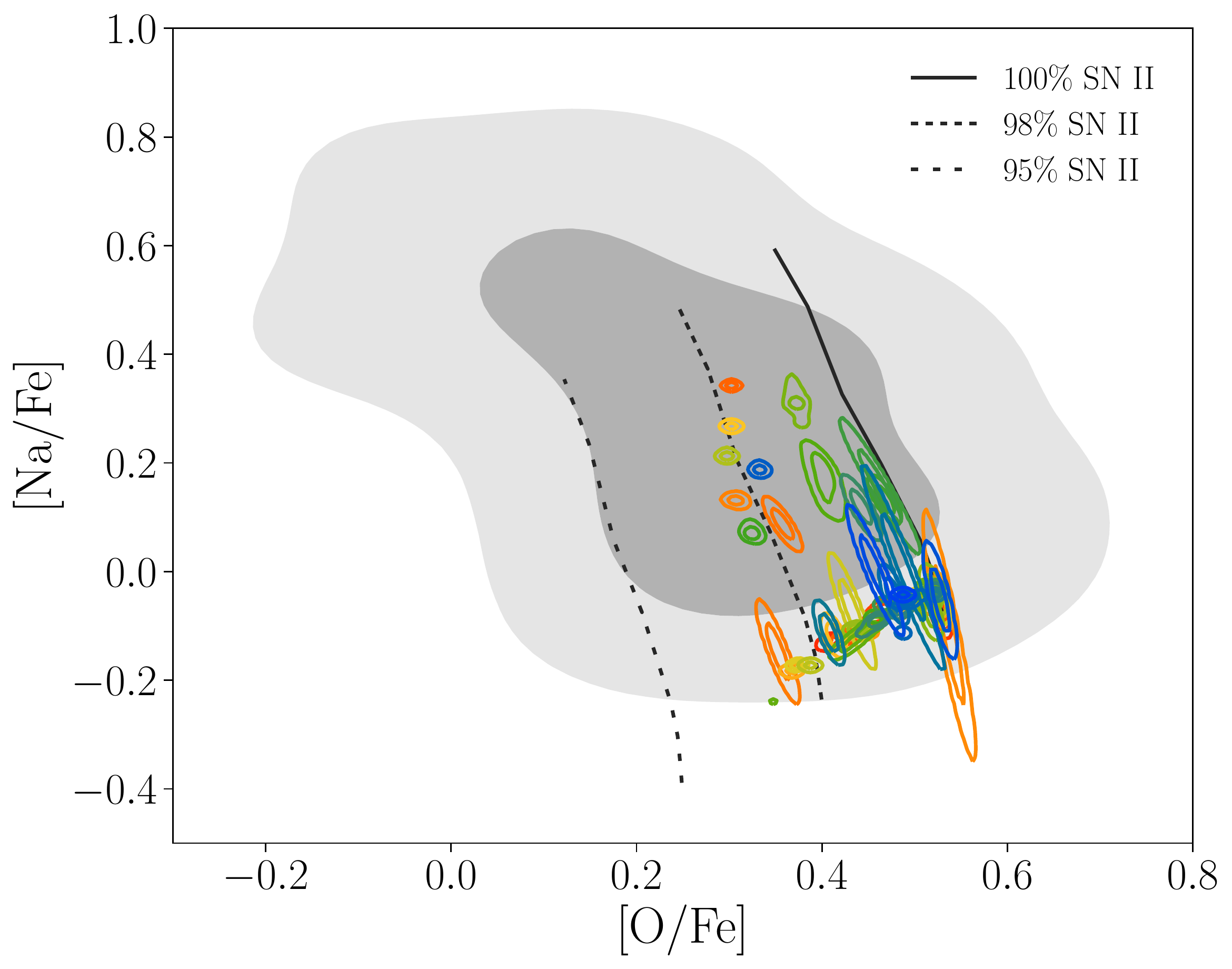}
    \caption{Na-O anticorrelation for model nuclear regions (small colored contours) vs. observed globular cluster stars (shaded contours) from \citet{Carretta_09_data}. The color of the lines simply serves to distinguish different model nuclear regions. Black lines show the elemental ratios of the SN yields, for different percentage contributions of SN type II. [Fe/H] varies along each line from -3 to 0.}
    \label{fig:na_o}
\end{figure}

\begin{figure}
	\includegraphics[width=0.48\textwidth]{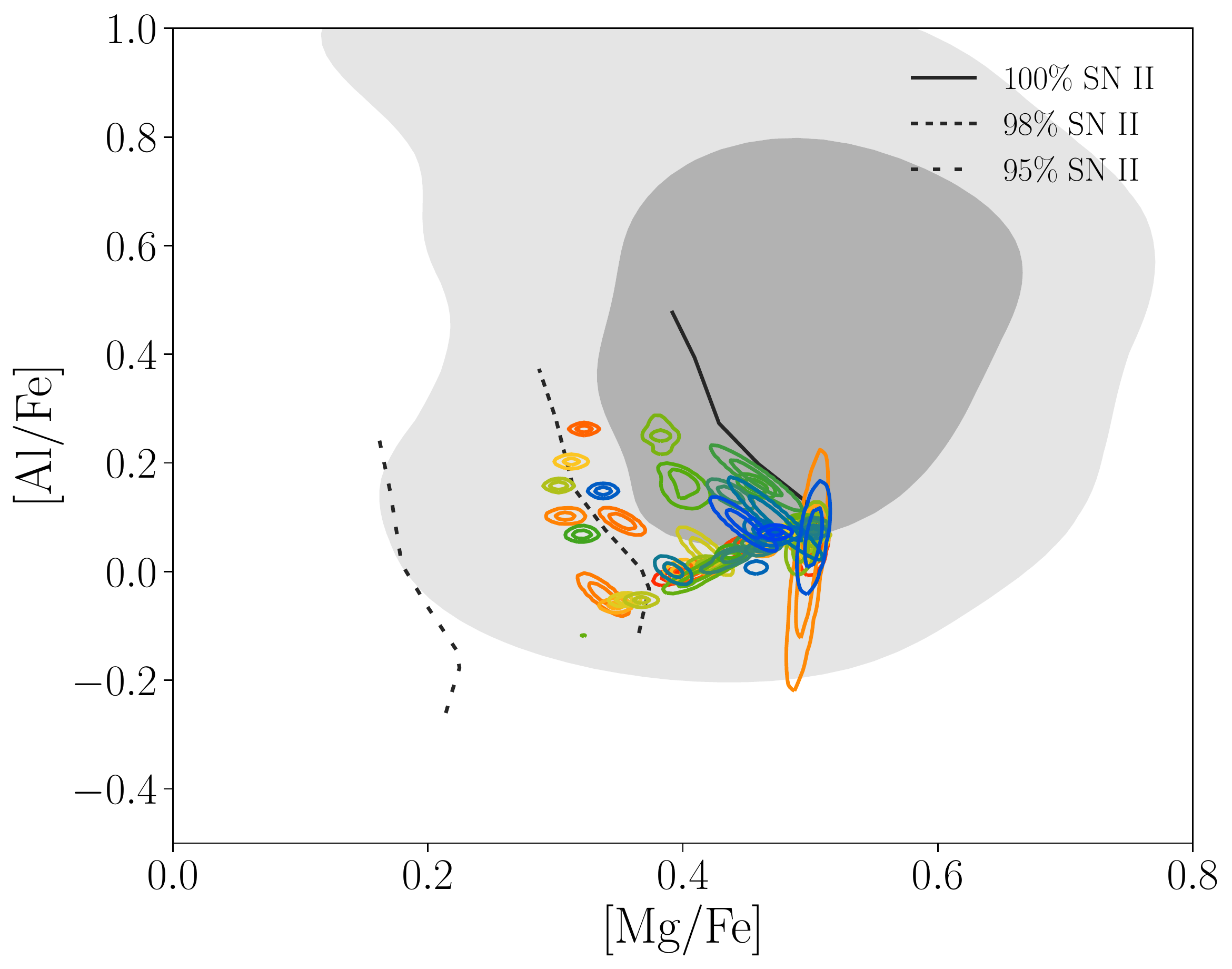}
    \caption{Same as \autoref{fig:na_o}, but for Al-Mg anticorrelation.}
    \label{fig:mg_al}
\end{figure}

Additionally, we examine the elemental anti-correlations (Na-O and Al-Mg) that are key features of the multiple population phenomena in globular clusters. We compute the [Na/Fe] and [O/Fe] ratios for each star particle. Since the underlying metallicity dispersion is responsible for the elemental dispersion of a star particle, the elemental dispersions will be correlated. A given metallicity spread will cause a spread along a certain direction in the [Na/Fe]-[O/Fe] plane. We smooth each star particle by a Gaussian  whose width along this direction will reproduce the elemental spreads (see \autoref{eq:elt_disp}). For visual clarity, a lower threshold on this width was chosen to be 0.005 dex, which was also used as the width perpendicular to the direction of the spread. We mass-weight the contributions of all star particles in a cluster, then plot contours that enclose 50\% and 90\% of the mass for each cluster. \autoref{fig:na_o} shows the results of this for the Na-O anticorrelation, while  \autoref{fig:mg_al} shows the Mg-Al anti-correlation. Black lines show the tracks that objects would follow if their metallicity came from a given percentage of Type II supernovae, for metallicities $-3 < \feh < 0$. Nuclear regions consisting of a single particle will exhibit spreads along these lines, while regions with multiple particles can have spreads in any direction due to potential variations between  particles. It is apparent that nearly all of our simulated nuclear regions do not have a significant spread in any elemental abundance, unlike stars in globular clusters. It is also apparent that the yield lines are not capable of producing the full range of elemental variations, especially for Al, even though they include metallicities up to solar.

As seen in \autoref{eq:elt_disp}, elements whose yields change more with metallicity will have larger spreads. Na and Al yields change more than O and Mg (see \autoref{fig:sn_z}), explaining their larger abundance spreads in these figures. 

\begin{table*}[t]
\centering
\caption{Elements and Yields}
\begin{tabular}{p{0.4in}p{1.7in}p{1.7in}p{2.7in}}
\tableline \\[-1mm]
Element & Importance and Measurements & Production Source & Reliability of Theoretical Yields 
  \\[2mm] \tableline\\[-1mm]
O & Depleted in second-generation (SG) stars as part of Na-O anticorrelation (R15, C09).
  & Massive stars (A17). Yield not affected by explosive nucleosynthesis (P16).  
    & Good agreement between galactic chemical evolution (GCE) models and observations. Predictions may vary by as much as 0.6~dex at lower metallicities, but less than 0.2~dex at solar metallicity (R10, A17).
  \\[1mm] \tableline \\[-1mm]
Na & Enriched in SG stars as part of Na-O anticorrelation (R15, C09).  
  & Massive stars (A17). Mainly comes from hydrostatic carbon burning, and is consumed in explosions (P16).
  & Strong metallicity dependence that does not match observations well at low metallicity (R10, A17). Models have large ($>0.5$~dex) spread at low metallicity (R10).
  \\[1mm] \tableline \\[-1mm]
Mg & Depleted in SG stars as part of Mg-Al anticorrelation (G12, FT17).
  & Massive stars (A17). Created both in the pre-SN and explosive nucleosynthesis phases (R10, P16). 
  & Matches observations well (R10, A17). Relatively low spread between models ($\sim 0.3$~dex) for all but the lowest metallicities (W09, R10).
  \\[1mm] \tableline \\[-1mm]
Al & Enriched in SG stars as part of Mg-Al anticorrelation (G12, FT17). 
  & Massive stars (A17). Created both in the pre-SN and explosive nucleosynthesis phases (P16).
  & Moderate metallicity dependence that does not match observations, but it closer than the Na yield. Spread in models is smaller than for Na, but larger than for O and Mg (R10, A17). 
  \\[1mm] \tableline \\[-1mm]
Fe & Some very massive "anomalous" clusters have spread in Fe: $\omega$~Cen, M54, Terzan~5 (M15).
   & SN type Ia and type II (A17).
   & Models are consistent with each other (W09). The adopted mass cut and details of the explosion are the primary drivers of discrepancy (W09, R10).
  \\[1mm] \tableline \\[-1mm]
\end{tabular} 
  Citations are abbreviated as follows -- A17: \citet{Andrews17}, C09: \citet{Carretta09_Na_O}, FT17: \citet{Fernandez-Trincado17}, G12: \citet{Gratton12}, M15: \citet{Marino15},  P16: \citet{PignatariNuGrid}, R10: \citet{Romano10}, R15: \citet{Renzini15}, W09: \citet{Wiersma09}.
  \\[4mm]
  \label{tab:gc_relevance}
\end{table*}

\section{Discussion}

The results presented above test the hypothesis that the abundances of nuclear star clusters are consistent with the observed trends in massive globular clusters. \rev{Here we discuss these results and the possible presence of dark matter within the NSCs.}

\subsection{Elemental Abundances}

As shown in \autoref{ssec:correlations}, we are unable to reproduce the full spread in light element abundances that are present in globular clusters. For [O/Fe] and [Mg/Fe], the one-sided spread is less than 0.02~dex for all clusters, while for [Na/Fe] and [Al/Fe] the spread is 0.1~dex at most. The observed one-sided spreads are significantly larger: about 0.2~dex for [O/Fe] and [Na/Fe], 0.1~dex for [Mg/Fe], and 0.3~dex for [Al/Fe]. The higher spread in Na and Al in our clusters is due to the stronger variation in the yields of these elements with metallicity (\autoref{fig:sn_z}). Even with this large variation, even the yields themselves are not able to produce as much Al as is present in enriched second-generation globular cluster stars.

The spread we calculate includes both the internal spread within a star particle due to a metallicity variation in the accreted gas during its formation, as well as the spread among different star particles within a given nuclear region. Even with both of these sources, the elemental spreads are too small. Our simulations would be able to reproduce the distinct stellar populations by having multiple star particles with differently enriched material, but we do not see this. Many nuclear regions consist of only one star particle, but those that have multiple do not show the necessary variations. In addition, a substantial age spread cannot be responsible for the elemental spreads, as none of our nuclear regions reach the required light element spread, regardless of their formation timescale. 

\rev{As described in \autoref{ssec:mass_size}, the sizes of our nuclear regions are an acceptable description of the central component of the central density component, even though they are typically larger than observed NSCs. This indicates that the particles that are included within each NSC are a reasonable selection. The main conclusion of our calculation is that the extended star formation histories, without internal pollution, are insufficient to reproduce the observed abundance trends in globular clusters. This conclusion would only be strengthened if our selection of NSC particles is an overestimate. If we reduce the sizes and masses of model NSCs, their abundance spread would only decrease.}

This indicates that our simulations do not contain the enrichment sources that are the true cause of the spreads. As supernovae are the only enrichment source we include, these results provide support to the idea that the polluter responsible for globular cluster abundances must lie within the clusters during their formation epoch. 
\rev{A number of models have been proposed to explain these light element abundance variations. Most scenarios call for multiple generations of star formation within the globular cluster's formation epoch, with enriched material from a first generation of stars seeding a second metal enriched generation.}
Possible versions suggested in the literature include supermassive stars \citep{Denissenkov_Hartwick_13}, rotating massive stars \citep{Decressin07,Krause13}, AGB stars \citep{Cottrell_daCosta_81,Renzini15}, and interacting binaries \citep{deMink09}. However, all of these current scenarios have some shortcomings in matching the full set of available observations, as discussed by \citet{bastian_lardo18}. 
\rev{These models fail to produce correct elemental abundances of the globular clusters, both for the elements listed above and Helium. Additionally, to reproduce the large fraction of enriched stars in a globular cluster, most models require huge amounts of mass loss (of only first generation stars), which does not appear to be realistic \citep{Larsen_Strader_Brodie_12,Cabrera-ziri_15}.}

\revb{One important aspect of our simulations is that they only reach $z\sim 1.5$. As later star formation would make the masses, metallicity spreads, and age spreads increase, the quantities we report for those properties are best interpreted as lower limits. This complicates the comparison of our NSCs to observed NSCs and GCs, which are all at $z\approx 0$. For example, the later bursts of star formation in Terzan 5 and M54 would not be present in our simulation.}

\revb{With this caveat in mind, we will discuss the predicted iron spread in model NSCs}. The majority of clusters have a spread less than 0.1 dex, which is roughly the limit of the observational sensitivity \citep[e.g.,][]{gratton_etal04, willman_strader12}. However, there are some clusters with $\feh$ spreads larger than this that could be the progenitors of the iron-complex GCs. These model NSCs are very metal-poor, with the metallicity distribution dominated by internal dispersion. Our simulations include also two nuclear regions that have a longer formation timescale and a $\feh$ spread dominated by variation among the star particles. These regions may be analogous to clusters like $\omega$~Cen or M54, which may share a similar origin \citep{Carretta_10_omega_cen_M54}. \rev{Our objects formed on a timescale consistent with that of $\omega$~Cen, which appears to have an age spread of \revb{at least} 500 Myr \citep{tailo_etal16}}. The metallicity distribution of the object with the largest spread between particles has a central peak at a metallicity consistent with the peak of the metallicity distribution of $\omega$~Cen and M54, but the tail extends to lower metallicities rather than higher. Our objects have lower mass than the observed clusters as well. These quantitative differences need to be explored in future work, but our results lend support to the hypothesis that objects like $\omega$~Cen could be made through tidal stripping of NSCs.

\rev{If this model is correct, it has implications for the location of the polluter responsible for the light element spreads of the massive GCs. If the NSC formed by in-situ formation \citep[e.g.][]{M_Hartmann_etal_2011}, the polluter must be present for each new burst of star formation. However, if the NSC formed by the inspiral of globular clusters \citep[e.g.][]{antonini12, gnedin_etal14, tsatsi17}, the presence of already existing light element spreads in each cluster would provide a natural explanation for the fact that light element spreads are found within each peak in the metallicity distribution of the iron-complex GCs. No polluter would be needed in the central region of the galaxy. Future observations may be able to test these hypotheses by looking for light element abundance spreads in NSCs. If not detected, both the model of massive GC formation discussed in this paper and the GC inspiral model of NSC formation would need to be discarded. If light element abundance variations are present in NSCs, these models would remain plausible.}

\subsection{Dark Matter}

The amount of dark matter contained within the extent of NSCs is important for their ability to represent progenitors of massive globular clusters. Even the largest globular clusters can have only dynamically sub-dominant amount of dark matter within their optical radii \citep[e.g.,][]{Conroy11, Ibata13}. Since NSCs form at the bottom of the dark matter potential well, they \rev{could} retain some of the dark matter even if the rest of the galaxy is tidally stripped.

We find that only one in every 11 NSCs contains any dark matter particles within $\Rnsc$. This does not necessarily imply that no dark matter is present. The mass and spatial resolution for dark matter is relatively low in our simulations and may prevent reliable estimate of the dark matter content within the compact clusters. The dark matter particle mass is about $10^6\Msun$, so that even one particle can be more massive than the whole NSC. Also, the spatial force resolution for dark matter is kept separate from the gas and star particle resolution, at $50-100\,$pc, so that the discreteness of massive dark matter particles does not affect the dynamics of gas and stars. This resolution makes it difficult to put strict constraints on the amount of dark matter present, but we can still conclude that dark matter is dynamically sub-dominant within the model nuclear regions. If these central regions turned into an object resembling a massive globular cluster after being tidally stripped by the host galaxy, the remaining amount of dark matter would likely not violate observational constraints.

\section{Summary}

We have investigated the origin of stellar populations of nuclear star clusters in galaxy formation simulations. We analyzed the outputs of a suite of simulations at redshift $z \approx 1.5$ and identified nuclear regions of galaxies using a two-component fit of the stellar surface density profile. The range of inferred NSC masses is consistent with the observed nuclear regions of nearby galaxies, although the sizes are generally overestimated. We find that the shapes of model NSCs are moderately flattened, but not by rotation.

\rev{We examine two of the deviations from a simple stellar population present in massive globular clusters (age and metallicity spreads), and how they affect light element abundances. Age spreads are derived directly from the simulation outputs, while}
we calculated the abundances of several elements (Fe, O, Na, Mg, Al) by applying theoretical model yields to Type II and Ia supernova ejecta calculated in the simulations.

Our main results can be summarized as follows.

$\bullet$ The nuclear regions are systematically more metal-rich than their host galaxies. The average metallicities of galaxies in the simulations match the observed mass-metallicity relation for galaxies at $z\approx 1.5$.

$\bullet$ We find some nuclear regions with a large spread in Fe that could be progenitors of objects like M54 or $\omega$ Cen.

$\bullet$ The predicted spread of light element abundances in NSCs is significantly smaller than that observed in globular clusters\rev{, even in clusters with a large age or $\feh$ spread.}

$\bullet$ We find no clear dependence of these trends on the local efficiency of star formation $\epsff$ used in our simulations.

$\bullet$ Our nuclear regions do not contain significant quantities of dark matter.

\rev{Our results show that NSCs can plausibly be the progenitors of the massive iron-complex globular clusters. However, these metallicity spreads cannot contribute significantly to the observed light element spreads.} 
The observed abundance spread must involve additional sources, internal to the clusters.

\acknowledgments 
We thank the anonymous referee for constructive comments that improved the paper. We also thank Eric Bell, Monica Valluri, Kohei Hattori, and Ian Roederer for helpful discussions. 
This work was supported in part by NSF through grant 1412144.

\makeatletter\@chicagotrue\makeatother

\bibliographystyle{yahapj}
\bibliography{bibfile,gc}

\begin{thebibliography}{}
\providecommand\natexlab[1]{#1}
\providecommand\JournalTitle[1]{#1}

\bibitem[{{Andrews} {et~al.}(2017){Andrews}, {Weinberg}, {Sch{\"o}nrich}, \&
  {Johnson}}]{Andrews17}
{Andrews}, B.~H., {Weinberg}, D.~H., {Sch{\"o}nrich}, R., \& {Johnson}, J.~A.
  2017,
  \href{http://dx.doi.org/10.3847/1538-4357/835/2/224}{\JournalTitle{\apj},
  835, 224}

\bibitem[{{Antonini} {et~al.}(2012){Antonini}, {Capuzzo-Dolcetta},
  {Mastrobuono- Battisti}, \& {Merritt}}]{antonini12}
{Antonini}, F., {Capuzzo-Dolcetta}, R., {Mastrobuono- Battisti}, A., \&
  {Merritt}, D. 2012,
  \href{http://dx.doi.org/10.1088/0004-637X/750/2/111}{\JournalTitle{\apj},
  750, 111}

\bibitem[{{Bastian} \& {Lardo}(2018)}]{bastian_lardo18}
{Bastian}, N., \& {Lardo}, C. 2018, \JournalTitle{ArXiv e-prints},
  \href{http://arxiv.org/abs/1712.01286}{{\sffamily arXiv:1712.01286
  [astro-ph.SR]}}

\bibitem[{{Behroozi} {et~al.}(2013){Behroozi}, {Wechsler}, \& {Wu}}]{rockstar}
{Behroozi}, P.~S., {Wechsler}, R.~H., \& {Wu}, H.-Y. 2013,
  \href{http://dx.doi.org/10.1088/0004-637X/762/2/109}{\JournalTitle{\apj},
  762, 109}

\bibitem[{{B{\"o}ker}(2008)}]{Boker08}
{B{\"o}ker}, T. 2008,
  \href{http://dx.doi.org/10.1086/527033}{\JournalTitle{\apjl}, 672, L111}

\bibitem[{{Cabrera-Ziri} {et~al.}(2015){Cabrera-Ziri}, {Bastian}, {Longmore},
  {Brogan}, {Hollyhead}, {Larsen}, {Whitmore}, {Johnson}, {Chandar}, {Henshaw},
  {Davies}, \& {Hibbard}}]{Cabrera-ziri_15}
{Cabrera-Ziri}, I., {Bastian}, N., {Longmore}, S.~N., {et~al.} 2015,
  \href{http://dx.doi.org/10.1093/mnras/stv163}{\JournalTitle{\mnras}, 448,
  2224}

\bibitem[{{Carretta} {et~al.}(2009{\natexlab{a}}){Carretta}, {Bragaglia},
  {Gratton}, \& {Lucatello}}]{Carretta_09_data}
{Carretta}, E., {Bragaglia}, A., {Gratton}, R., \& {Lucatello}, S.
  2009{\natexlab{a}},
  \href{http://dx.doi.org/10.1051/0004-6361/200912097}{\JournalTitle{\aap},
  505, 139}

\bibitem[{{Carretta} {et~al.}(2009{\natexlab{b}}){Carretta}, {Bragaglia},
  {Gratton}, {Lucatello}, {Catanzaro}, {Leone}, {Bellazzini}, {Claudi},
  {D'Orazi}, {Momany}, {Ortolani}, {Pancino}, {Piotto}, {Recio-Blanco}, \&
  {Sabbi}}]{Carretta09_Na_O}
{Carretta}, E., {Bragaglia}, A., {Gratton}, R.~G., {et~al.} 2009{\natexlab{b}},
  \href{http://dx.doi.org/10.1051/0004-6361/200912096}{\JournalTitle{\aap},
  505, 117}

\bibitem[{{Carretta} {et~al.}(2010{\natexlab{a}}){Carretta}, {Bragaglia},
  {Gratton}, {Lucatello}, {Bellazzini}, {Catanzaro}, {Leone}, {Momany},
  {Piotto}, \& {D'Orazi}}]{Carretta_etal_10_m54}
---. 2010{\natexlab{a}},
  \href{http://dx.doi.org/10.1051/0004-6361/201014924}{\JournalTitle{\aap},
  520, A95}

\bibitem[{{Carretta} {et~al.}(2010{\natexlab{b}}){Carretta}, {Bragaglia},
  {Gratton}, {Lucatello}, {Bellazzini}, {Catanzaro}, {Leone}, {Momany},
  {Piotto}, \& {D'Orazi}}]{Carretta_10_omega_cen_M54}
---. 2010{\natexlab{b}},
  \href{http://dx.doi.org/10.1088/2041-8205/714/1/L7}{\JournalTitle{\apj}, 714,
  L7}

\bibitem[{{Conroy} {et~al.}(2011){Conroy}, {Loeb}, \& {Spergel}}]{Conroy11}
{Conroy}, C., {Loeb}, A., \& {Spergel}, D.~N. 2011,
  \href{http://dx.doi.org/10.1088/0004-637X/741/2/72}{\JournalTitle{\apj}, 741,
  72}

\bibitem[{{Cottrell} \& {Da Costa}(1981)}]{Cottrell_daCosta_81}
{Cottrell}, P.~L., \& {Da Costa}, G.~S. 1981,
  \href{http://dx.doi.org/10.1086/183527}{\JournalTitle{\apjl}, 245, L79}

\bibitem[{{de Mink} {et~al.}(2009){de Mink}, {Pols}, {Langer}, \&
  {Izzard}}]{deMink09}
{de Mink}, S.~E., {Pols}, O.~R., {Langer}, N., \& {Izzard}, R.~G. 2009,
  \href{http://dx.doi.org/10.1051/0004-6361/200913205}{\JournalTitle{\aap},
  507, L1}

\bibitem[{{Decressin} {et~al.}(2007){Decressin}, {Meynet}, {Charbonnel},
  {Prantzos}, \& {Ekstr{\"o}m}}]{Decressin07}
{Decressin}, T., {Meynet}, G., {Charbonnel}, C., {Prantzos}, N., \&
  {Ekstr{\"o}m}, S. 2007,
  \href{http://dx.doi.org/10.1051/0004-6361:20066013}{\JournalTitle{\aap}, 464,
  1029}

\bibitem[{{Dejonghe}(1987)}]{Dejonghe87}
{Dejonghe}, H. 1987,
  \href{http://dx.doi.org/10.1093/mnras/224.1.13}{\JournalTitle{\mnras}, 224,
  13}

\bibitem[{{Denissenkov} \& {Hartwick}(2014)}]{Denissenkov_Hartwick_13}
{Denissenkov}, P.~A., \& {Hartwick}, F.~D.~A. 2014,
  \href{http://dx.doi.org/10.1093/mnrasl/slt133}{\JournalTitle{\mnras}, 437,
  L21}

\bibitem[{{Do} {et~al.}(2015){Do}, {Kerzendorf}, {Winsor}, {St{\o}stad},
  {Morris}, {Lu}, \& {Ghez}}]{Do15}
{Do}, T., {Kerzendorf}, W., {Winsor}, N., {et~al.} 2015,
  \href{http://dx.doi.org/10.1088/0004-637X/809/2/143}{\JournalTitle{\apj},
  809, 143}

\bibitem[{{Feldmeier-Krause} {et~al.}(2017{\natexlab{a}}){Feldmeier-Krause},
  {Kerzendorf}, {Neumayer}, {Sch{\"o}del}, {Nogueras-Lara}, {Do}, {de Zeeuw},
  \& {Kuntschner}}]{FK17_metallicity}
{Feldmeier-Krause}, A., {Kerzendorf}, W., {Neumayer}, N., {et~al.}
  2017{\natexlab{a}},
  \href{http://dx.doi.org/10.1093/mnras/stw2339}{\JournalTitle{\mnras}, 464,
  194}

\bibitem[{{Feldmeier-Krause} {et~al.}(2017{\natexlab{b}}){Feldmeier-Krause},
  {Zhu}, {Neumayer}, {van de Ven}, {de Zeeuw}, \& {Sch{\"o}del}}]{FK17_orbits}
{Feldmeier-Krause}, A., {Zhu}, L., {Neumayer}, N., {et~al.} 2017{\natexlab{b}},
  \href{http://dx.doi.org/10.1093/mnras/stw3377}{\JournalTitle{\mnras}, 466,
  4040}

\bibitem[{{Fern{\'a}ndez-Trincado} {et~al.}(2017){Fern{\'a}ndez-Trincado},
  {Zamora}, {Garc{\'{\i}}a-Hern{\'a}ndez}, {Souto}, {Dell'Agli}, {Schiavon},
  {Geisler}, {Tang}, {Villanova}, {Hasselquist}, {Mennickent}, {Cunha},
  {Shetrone}, {Allende Prieto}, {Vieira}, {Zasowski}, {Sobeck}, {Hayes},
  {Majewski}, {Placco}, {Beers}, {Schleicher}, {Robin}, {M{\'e}sz{\'a}ros},
  {Masseron}, {Garc{\'{\i}}a P{\'e}rez}, {Anders}, {Meza}, {Alves-Brito},
  {Carrera}, {Minniti}, {Lane}, {Fern{\'a}ndez-Alvar}, {Moreno}, {Pichardo},
  {P{\'e}rez-Villegas}, {Schultheis}, {Roman-Lopes}, {Fuentes}, {Nitschelm},
  {Harding}, {Bizyaev}, {Pan}, {Oravetz}, {Simmons}, {Ivans},
  {Blanco-Cuaresma}, {Hern{\'a}ndez}, {Alonso-Garc{\'{\i}}a}, {Valenzuela}, \&
  {Chanam{\'e}}}]{Fernandez-Trincado17}
{Fern{\'a}ndez-Trincado}, J.~G., {Zamora}, O., {Garc{\'{\i}}a-Hern{\'a}ndez},
  D.~A., {et~al.} 2017,
  \href{http://dx.doi.org/10.3847/2041-8213/aa8032}{\JournalTitle{\apjl}, 846,
  L2}

\bibitem[{{Ferrarese} {et~al.}(2006){Ferrarese}, {C{\^o}t{\'e}}, {Dalla
  Bont{\`a}}, {Peng}, {Merritt}, {Jord{\'a}n}, {Blakeslee}, {Ha{\c s}egan},
  {Mei}, {Piatek}, {Tonry}, \& {West}}]{Ferrarese06}
{Ferrarese}, L., {C{\^o}t{\'e}}, P., {Dalla Bont{\`a}}, E., {et~al.} 2006,
  \href{http://dx.doi.org/10.1086/505388}{\JournalTitle{\apjl}, 644, L21}

\bibitem[{{Ferraro} {et~al.}(2016){Ferraro}, {Massari}, {Dalessandro},
  {Lanzoni}, {Origlia}, {Rich}, \& {Mucciarelli}}]{Ferraro16}
{Ferraro}, F.~R., {Massari}, D., {Dalessandro}, E., {et~al.} 2016,
  \href{http://dx.doi.org/10.3847/0004-637X/828/2/75}{\JournalTitle{\apj}, 828,
  75}

\bibitem[{{Freeman}(1993)}]{Freeman93}
{Freeman}, K.~C. 1993, in Astronomical Society of the Pacific Conference
  Series, Vol.~48, The Globular Cluster-Galaxy Connection, ed. G.~H. {Smith} \&
  J.~P. {Brodie}, 608

\bibitem[{{Georgiev} {et~al.}(2016){Georgiev}, {B{\"o}ker}, {Leigh},
  {L{\"u}tzgendorf}, \& {Neumayer}}]{Georgiev16}
{Georgiev}, I.~Y., {B{\"o}ker}, T., {Leigh}, N., {L{\"u}tzgendorf}, N., \&
  {Neumayer}, N. 2016,
  \href{http://dx.doi.org/10.1093/mnras/stw093}{\JournalTitle{\mnras}, 457,
  2122}

\bibitem[{{Gnedin}(2014)}]{n_gnedin_14_reionization}
{Gnedin}, N.~Y. 2014,
  \href{http://dx.doi.org/10.1088/0004-637X/793/1/29}{\JournalTitle{\apj}, 793}

\bibitem[{{Gnedin} \& {Abel}(2001)}]{ngnedin_abel01}
{Gnedin}, N.~Y., \& {Abel}, T. 2001,
  \href{http://dx.doi.org/10.1016/S1384-1076(01)00068-9}{\JournalTitle{\na}, 6,
  437}

\bibitem[{{Gnedin} \& {Kravtsov}(2011)}]{ngnedin_kravtsov11}
{Gnedin}, N.~Y., \& {Kravtsov}, A.~V. 2011,
  \href{http://dx.doi.org/10.1088/0004-637X/728/2/88}{\JournalTitle{\apj}, 728,
  88}

\bibitem[{{Gnedin} {et~al.}(2014){Gnedin}, {Ostriker}, \&
  {Tremaine}}]{gnedin_etal14}
{Gnedin}, O.~Y., {Ostriker}, J.~P., \& {Tremaine}, S. 2014,
  \href{http://dx.doi.org/10.1088/0004-637X/785/1/71}{\JournalTitle{\apj}, 785,
  71}

\bibitem[{{Gratton} {et~al.}(2004){Gratton}, {Sneden}, \&
  {Carretta}}]{gratton_etal04}
{Gratton}, R., {Sneden}, C., \& {Carretta}, E. 2004,
  \href{http://dx.doi.org/10.1146/annurev.astro.42.053102.133945}{\JournalTitle{\araa},
  42, 385}

\bibitem[{{Gratton} {et~al.}(2012){Gratton}, {Carretta}, \&
  {Bragaglia}}]{Gratton12}
{Gratton}, R.~G., {Carretta}, E., \& {Bragaglia}, A. 2012,
  \href{http://dx.doi.org/10.1007/s00159-012-0050-3}{\JournalTitle{\aapr}, 20,
  50}

\bibitem[{{Grevesse} \& {Sauval}(1998)}]{Grevesse_Sauval_98}
{Grevesse}, N., \& {Sauval}, A.~J. 1998,
  \href{http://dx.doi.org/10.1023/A:1005161325181}{\JournalTitle{\ssr}, 85,
  161}

\bibitem[{{Haardt} \& {Madau}(2001)}]{haardt_madau_01}
{Haardt}, F., \& {Madau}, P. 2001, in Clusters of galaxies and the high
  redshift universe observed in X-rays, Recent results of XMM-Newton and
  Chandra, XXXVIth Rencontres de Moriond , XXIst Moriond Astrophysics Meeting,
  March 10-17, 2001 Savoie, France. Edited by D.M. Neumann \& J.T.T. Van. <A
  href="http://moriond.in2p3.fr">http://moriond.in2p3.fr</A>, id. 64

\bibitem[{{Hartmann} {et~al.}(2011){Hartmann}, {Debattista}, {Seth},
  {Cappellari}, \& {Quinn}}]{M_Hartmann_etal_2011}
{Hartmann}, M., {Debattista}, V.~P., {Seth}, A., {Cappellari}, M., \& {Quinn},
  T.~R. 2011,
  \href{http://dx.doi.org/10.1111/j.1365-2966.2011.19659.x}{\JournalTitle{\mnras},
  418, 2697}

\bibitem[{{Ibata} {et~al.}(2013){Ibata}, {Nipoti}, {Sollima}, {Bellazzini},
  {Chapman}, \& {Dalessandro}}]{Ibata13}
{Ibata}, R., {Nipoti}, C., {Sollima}, A., {et~al.} 2013,
  \href{http://dx.doi.org/10.1093/mnras/sts302}{\JournalTitle{\mnras}, 428,
  3648}

\bibitem[{{Iwamoto} {et~al.}(1999){Iwamoto}, {Brachwitz}, {Nomoto},
  {Kishimoto}, {Umeda}, {Hix}, \& {Thielemann}}]{Iwamoto99}
{Iwamoto}, K., {Brachwitz}, F., {Nomoto}, K., {et~al.} 1999,
  \href{http://dx.doi.org/10.1086/313278}{\JournalTitle{\apjs}, 125, 439}

\bibitem[{{Johnson} {et~al.}(2009){Johnson}, {Pilachowski}, {Michael Rich}, \&
  {Fulbright}}]{johnson_etal09}
{Johnson}, C.~I., {Pilachowski}, C.~A., {Michael Rich}, R., \& {Fulbright},
  J.~P. 2009,
  \href{http://dx.doi.org/10.1088/0004-637X/698/2/2048}{\JournalTitle{\apj},
  698, 2048}

\bibitem[{{Joo} \& {Lee}(2013)}]{Joo_Lee_13}
{Joo}, S.-J., \& {Lee}, Y.-W. 2013,
  \href{http://dx.doi.org/10.1088/0004-637X/762/1/36}{\JournalTitle{\apj}, 762}

\bibitem[{{Kirby} {et~al.}(2013){Kirby}, {Cohen}, {Guhathakurta}, {Cheng},
  {Bullock}, \& {Gallazzi}}]{Kirby13}
{Kirby}, E.~N., {Cohen}, J.~G., {Guhathakurta}, P., {et~al.} 2013,
  \href{http://dx.doi.org/10.1088/0004-637X/779/2/102}{\JournalTitle{\apj},
  779, 102}

\bibitem[{{Krause} {et~al.}(2013){Krause}, {Charbonnel}, {Decressin}, {Meynet},
  \& {Prantzos}}]{Krause13}
{Krause}, M., {Charbonnel}, C., {Decressin}, T., {Meynet}, G., \& {Prantzos},
  N. 2013,
  \href{http://dx.doi.org/10.1051/0004-6361/201220694}{\JournalTitle{\aap},
  552, A121}

\bibitem[{{Kravtsov}(1999)}]{kravtsov99}
{Kravtsov}, A.~V. 1999, PhD thesis, New Mexico State University

\bibitem[{{Kravtsov}(2003)}]{kravtsov03}
---. 2003, \JournalTitle{\apjl}, 590, L1

\bibitem[{Kravtsov {et~al.}(1997)Kravtsov, Klypin, \&
  Khokhlov}]{kravtsov_etal97}
Kravtsov, A.~V., Klypin, A.~A., \& Khokhlov, A.~M. 1997, \JournalTitle{\apjs},
  111, 73

\bibitem[{{Kroupa}(2001)}]{Kroupa01}
{Kroupa}, P. 2001,
  \href{http://dx.doi.org/10.1046/j.1365-8711.2001.04022.x}{\JournalTitle{\mnras},
  322, 231}

\bibitem[{{Larsen} {et~al.}(2012){Larsen}, {Strader}, \&
  {Brodie}}]{Larsen_Strader_Brodie_12}
{Larsen}, S.~S., {Strader}, J., \& {Brodie}, J.~P. 2012,
  \href{http://dx.doi.org/10.1051/0004-6361/201219897}{\JournalTitle{\aap},
  544}

\bibitem[{{Leigh} {et~al.}(2012){Leigh}, {B{\"o}ker}, \& {Knigge}}]{Leigh12}
{Leigh}, N., {B{\"o}ker}, T., \& {Knigge}, C. 2012,
  \href{http://dx.doi.org/10.1111/j.1365-2966.2012.21365.x}{\JournalTitle{\mnras},
  424, 2130}

\bibitem[{{Li} {et~al.}(2018){Li}, {Gnedin}, \& {Gnedin}}]{li_etal18}
{Li}, H., {Gnedin}, O.~Y., \& {Gnedin}, N.~Y. 2018, \JournalTitle{\apj,
  submitted; arXiv:1712.01219},
  \href{http://arxiv.org/abs/1712.01219}{{\sffamily arXiv:1712.01219}}

\bibitem[{{Li} {et~al.}(2017){Li}, {Gnedin}, {Gnedin}, {Meng}, {Semenov}, \&
  {Kravtsov}}]{li_etal17}
{Li}, H., {Gnedin}, O.~Y., {Gnedin}, N.~Y., {et~al.} 2017,
  \href{http://dx.doi.org/10.3847/1538-4357/834/1/69}{\JournalTitle{\apj}, 834,
  69}

\bibitem[{{Mannucci} {et~al.}(2009){Mannucci}, {Cresci}, {Maiolino}, {Marconi},
  {Pastorini}, {Pozzetti}, {Gnerucci}, {Risaliti}, {Schneider}, {Lehnert}, \&
  {Salvati}}]{Mannucci09}
{Mannucci}, F., {Cresci}, G., {Maiolino}, R., {et~al.} 2009,
  \href{http://dx.doi.org/10.1111/j.1365-2966.2009.15185.x}{\JournalTitle{\mnras},
  398, 1915}

\bibitem[{{Marino} {et~al.}(2011){Marino}, {Milone}, {Piotto}, {Villanova},
  {Gratton}, {D'Antona}, {Anderson}, {Bedin}, {Bellini}, {Cassisi}, {Geisler},
  {Renzini}, \& {Zoccali}}]{Marino11}
{Marino}, A.~F., {Milone}, A.~P., {Piotto}, G., {et~al.} 2011,
  \href{http://dx.doi.org/10.1088/0004-637X/731/1/64}{\JournalTitle{\apj}, 731,
  64}

\bibitem[{{Marino} {et~al.}(2015){Marino}, {Milone}, {Karakas}, {Casagrande},
  {Yong}, {Shingles}, {Da Costa}, {Norris}, {Stetson}, {Lind}, {Asplund},
  {Collet}, {Jerjen}, {Sbordone}, {Aparicio}, \& {Cassisi}}]{Marino15}
{Marino}, A.~F., {Milone}, A.~P., {Karakas}, A.~I., {et~al.} 2015,
  \href{http://dx.doi.org/10.1093/mnras/stv420}{\JournalTitle{\mnras}, 450,
  815}

\bibitem[{{Massari} {et~al.}(2014){Massari}, {Mucciarelli}, {Ferraro},
  {Origlia}, {Rich}, {Lanzoni}, {Dalessandro}, {Valenti}, {Ibata}, {Lovisi},
  {Bellazzini}, \& {Reitzel}}]{Massari_etal_10}
{Massari}, D., {Mucciarelli}, A., {Ferraro}, F.~R., {et~al.} 2014,
  \href{http://dx.doi.org/10.1088/0004-637X/795/1/22}{\JournalTitle{\apj}, 795,
  22}

\bibitem[{{Milone} {et~al.}(2017){Milone}, {Piotto}, {Renzini}, {Marino},
  {Bedin}, {Vesperini}, {D'Antona}, {Nardiello}, {Anderson}, {King}, {Yong},
  {Bellini}, {Aparicio}, {Barbuy}, {Brown}, {Cassisi}, {Ortolani}, {Salaris},
  {Sarajedini}, \& {van der Marel}}]{Milone17}
{Milone}, A.~P., {Piotto}, G., {Renzini}, A., {et~al.} 2017,
  \href{http://dx.doi.org/10.1093/mnras/stw2531}{\JournalTitle{\mnras}, 464,
  3636}

\bibitem[{{Mo} {et~al.}(2010){Mo}, {van den Bosch}, \&
  {White}}]{mo_vandenbosch_white_book}
{Mo}, H., {van den Bosch}, F.~C., \& {White}, S. 2010, {Galaxy Formation and
  Evolution}

\bibitem[{{Moustakas} {et~al.}(2010){Moustakas}, {Kennicutt}, {Tremonti},
  {Dale}, {Smith}, \& {Calzetti}}]{Moustakas_etal_10}
{Moustakas}, J., {Kennicutt}, Jr., R.~C., {Tremonti}, C.~A., {et~al.} 2010,
  \href{http://dx.doi.org/10.1088/0067-0049/190/2/233}{\JournalTitle{\apjs},
  190, 233}

\bibitem[{{Nomoto} {et~al.}(2006){Nomoto}, {Tominaga}, {Umeda}, {Kobayashi}, \&
  {Maeda}}]{Nomoto06}
{Nomoto}, K., {Tominaga}, N., {Umeda}, H., {Kobayashi}, C., \& {Maeda}, K.
  2006,
  \href{http://dx.doi.org/10.1016/j.nuclphysa.2006.05.008}{\JournalTitle{Nuclear
  Physics A}, 777, 424}

\bibitem[{{Pignatari} {et~al.}(2016){Pignatari}, {Herwig}, {Hirschi},
  {Bennett}, {Rockefeller}, {Fryer}, {Timmes}, {Ritter}, {Heger}, {Jones},
  {Battino}, {Dotter}, {Trappitsch}, {Diehl}, {Frischknecht}, {Hungerford},
  {Magkotsios}, {Travaglio}, \& {Young}}]{PignatariNuGrid}
{Pignatari}, M., {Herwig}, F., {Hirschi}, R., {et~al.} 2016,
  \href{http://dx.doi.org/10.3847/0067-0049/225/2/24}{\JournalTitle{\apjs},
  225, 24}

\bibitem[{{Renzini} {et~al.}(2015){Renzini}, {D'Antona}, {Cassisi}, {King},
  {Milone}, {Ventura}, {Anderson}, {Bedin}, {Bellini}, {Brown}, {Piotto}, {van
  der Marel}, {Barbuy}, {Dalessandro}, {Hidalgo}, {Marino}, {Ortolani},
  {Salaris}, \& {Sarajedini}}]{Renzini15}
{Renzini}, A., {D'Antona}, F., {Cassisi}, S., {et~al.} 2015,
  \href{http://dx.doi.org/10.1093/mnras/stv2268}{\JournalTitle{\mnras}, 454,
  4197}

\bibitem[{{Romano} {et~al.}(2010){Romano}, {Karakas}, {Tosi}, \&
  {Matteucci}}]{Romano10}
{Romano}, D., {Karakas}, A.~I., {Tosi}, M., \& {Matteucci}, F. 2010,
  \href{http://dx.doi.org/10.1051/0004-6361/201014483}{\JournalTitle{\aap},
  522, A32}

\bibitem[{{Rudd} {et~al.}(2008){Rudd}, {Zentner}, \& {Kravtsov}}]{rudd_etal08}
{Rudd}, D.~H., {Zentner}, A.~R., \& {Kravtsov}, A.~V. 2008,
  \href{http://dx.doi.org/10.1086/523836}{\JournalTitle{\apj}, 672, 19}

\bibitem[{{Seth} {et~al.}(2010){Seth}, {Cappellari}, {Neumayer}, {Caldwell},
  {Bastian}, {Olsen}, {Blum}, {Debattista}, {McDermid}, {Puzia}, \&
  {Stephens}}]{Seth10}
{Seth}, A.~C., {Cappellari}, M., {Neumayer}, N., {et~al.} 2010,
  \href{http://dx.doi.org/10.1088/0004-637X/714/1/713}{\JournalTitle{\apj},
  714, 713}

\bibitem[{{Siegel} {et~al.}(2007){Siegel}, {Dotter}, {Majewski}, {Sarajedini},
  {Chaboyer}, {Nidever}, {Anderson}, {Mar{\'\i}n-Franch}, {Rosenberg}, {Bedin},
  {Aparicio}, {King}, {Piotto}, \& {Reid}}]{Siegel07}
{Siegel}, M.~H., {Dotter}, A., {Majewski}, S.~R., {et~al.} 2007,
  \href{http://dx.doi.org/10.1086/522003}{\JournalTitle{\apj}, 667, L57}

\bibitem[{{Tailo} {et~al.}(2016){Tailo}, {Di Criscienzo}, {D'Antona}, {Caloi},
  \& {Ventura}}]{tailo_etal16}
{Tailo}, M., {Di Criscienzo}, M., {D'Antona}, F., {Caloi}, V., \& {Ventura}, P.
  2016, \href{http://dx.doi.org/10.1093/mnras/stw319}{\JournalTitle{\mnras},
  457, 4525}

\bibitem[{{Timmes} {et~al.}(1995){Timmes}, {Woosley}, \& {Weaver}}]{Timmes95}
{Timmes}, F.~X., {Woosley}, S.~E., \& {Weaver}, T.~A. 1995,
  \href{http://dx.doi.org/10.1086/192172}{\JournalTitle{\apjs}, 98, 617}

\bibitem[{{Tsatsi} {et~al.}(2017){Tsatsi}, {Mastrobuono-Battisti}, {van de
  Ven}, {Perets}, {Bianchini}, \& {Neumayer}}]{tsatsi17}
{Tsatsi}, A., {Mastrobuono-Battisti}, A., {van de Ven}, G., {et~al.} 2017,
  \href{http://dx.doi.org/10.1093/mnras/stw2593}{\JournalTitle{\mnras}, 464,
  3720}

\bibitem[{{van der Wel} {et~al.}(2014){van der Wel}, {Franx}, {van Dokkum},
  {Skelton}, {Momcheva}, {Whitaker}, {Brammer}, {Bell}, {Rix}, {Wuyts},
  {Ferguson}, {Holden}, {Barro}, {Koekemoer}, {Chang}, {McGrath},
  {H{\"a}ussler}, {Dekel}, {Behroozi}, {Fumagalli}, {Leja}, {Lundgren},
  {Maseda}, {Nelson}, {Wake}, {Patel}, {Labb{\'e}}, {Faber}, {Grogin}, \&
  {Kocevski}}]{VanDerWel14}
{van der Wel}, A., {Franx}, M., {van Dokkum}, P.~G., {et~al.} 2014,
  \href{http://dx.doi.org/10.1088/0004-637X/788/1/28}{\JournalTitle{\apj}, 788,
  28}

\bibitem[{{Villanova} {et~al.}(2014){Villanova}, {Geisler}, {Gratton}, \&
  {Cassisi}}]{Villanova_14_w_cen}
{Villanova}, S., {Geisler}, D., {Gratton}, R.~G., \& {Cassisi}, S. 2014,
  \href{http://dx.doi.org/10.1088/0004-637X/791/2/107}{\JournalTitle{\apj},
  791}

\bibitem[{{Wiersma} {et~al.}(2009){Wiersma}, {Schaye}, {Theuns}, {Dalla
  Vecchia}, \& {Tornatore}}]{Wiersma09}
{Wiersma}, R.~P.~C., {Schaye}, J., {Theuns}, T., {Dalla Vecchia}, C., \&
  {Tornatore}, L. 2009,
  \href{http://dx.doi.org/10.1111/j.1365-2966.2009.15331.x}{\JournalTitle{\mnras},
  399, 574}

\bibitem[{{Willman} \& {Strader}(2012)}]{willman_strader12}
{Willman}, B., \& {Strader}, J. 2012,
  \href{http://dx.doi.org/10.1088/0004-6256/144/3/76}{\JournalTitle{\aj}, 144,
  76}

\bibitem[{{Woosley} \& {Weaver}(1995)}]{WW95}
{Woosley}, S.~E., \& {Weaver}, T.~A. 1995,
  \href{http://dx.doi.org/10.1086/192237}{\JournalTitle{\apjs}, 101, 181}

\bibitem[{{Zemp} {et~al.}(2011){Zemp}, {Gnedin}, {Gnedin}, \&
  {Kravtsov}}]{Zemp11}
{Zemp}, M., {Gnedin}, O.~Y., {Gnedin}, N.~Y., \& {Kravtsov}, A.~V. 2011,
  \href{http://dx.doi.org/10.1088/0067-0049/197/2/30}{\JournalTitle{\apjs},
  197, 30}

\end{thebibliography}

\end{document}